\documentclass[twocolumn]{aastex63}
\usepackage{multirow}
\usepackage{lineno}
\usepackage{float}
\usepackage{pifont}
\newcommand{\cmark}{\ding{51}}%
\newcommand{\xmark}{\ding{55}}%
\newcommand{\pmark}{\ding{212}}%
\usepackage{graphicx}
\usepackage{lipsum}
\usepackage{rotating}

\accepted{ApJ}

\graphicspath{{./}{figures/}}

\begin{document}
\title{Constraints on the physical origin of large cavities in transition disks from multi-wavelength dust continuum emission}
\correspondingauthor{Anibal Sierra}
\email{anibalsierram@gmail.com}

\author[0000-0002-5991-8073]{Anibal Sierra}
\affiliation{Departamento de Astronom\'ia, Universidad de Chile, Camino El Observatorio 1515, Las Condes, Santiago, Chile}
\affiliation{Instituto de Astrofísica, Pontificia Universidad Católica de Chile, Av. Vicuña Mackenna 4860, 7820436 Macul, Santiago, Chile}
\affiliation{Mullard Space Science Laboratory, University College London, Holmbury St Mary, Dorking, Surrey RH5 6NT, UK}

\author[0000-0002-1199-9564]{Laura M. P\'erez}
\affiliation{Departamento de Astronom\'ia, Universidad de Chile, Camino El Observatorio 1515, Las Condes, Santiago, Chile}

\author[0009-0009-9182-3240]{Benjam\'in Sotomayor}
\affiliation{Departamento de Astronom\'ia, Universidad de Chile, Camino El Observatorio 1515, Las Condes, Santiago, Chile}

\author[0000-0002-7695-7605]{Myriam Benisty}
\affiliation{Univ. Grenoble Alpes, CNRS, IPAG, 38000 Grenoble, France}
\affiliation{Université Côte d'Azur, Observatoire de la Côte d'Azur, CNRS, Laboratoire Lagrange, France}

\author[0000-0002-7570-5596]{Claire J. Chandler}
\affiliation{National Radio Astronomy Observatory, P.O. Box O, Socorro, NM 87801, USA}

\author[0000-0003-2253-2270]{Sean Andrews}
\affiliation{Center for Astrophysics | Harvard \& Smithsonian, Cambridge, MA 02138, USA}

\author[0000-0003-2251-0602]{John Carpenter}
\affiliation{Joint ALMA Observatory, Avenida Alonso de Córdova 3107, Vitacura, Santiago, Chile}

\author[0000-0002-1493-300X]{Thomas Henning}
\affiliation{Max-Planck-Institut für Astronomie (MPIA), Königstuhl 17, 69117 Heidelberg, Germany}

\author[0000-0003-1859-3070]{Leonardo Testi}
\affiliation{Alma Mater Studiorum Università di Bologna, Dipartimento di Fisica e Astronomia (DIFA), Via Gobetti 93/2, 40129 Bologna, Italy}
\affiliation{INAF – Osservatorio Astrofisico di Arcetri, Largo E. Fermi 5, 50125 Firenze, Italy}

\author[0000-0001-8123-2943]{Luca Ricci}
\affiliation{Department of Physics and Astronomy, California State University Northridge, 18111 Nordhoff Street, Northridge, CA 91330, USA}

\author[0000-0003-1526-7587]{David Wilner}
\affiliation{Center for Astrophysics | Harvard \& Smithsonian, Cambridge, MA 02138, USA}

\begin{abstract}
The physical origin of the large cavities observed in transition disks is to date still unclear. Different physical mechanisms (e.g., a  companion, dead zones, enhanced grain growth) produce disk cavities of different depth, and the expected spatial distribution of gas and solids in each mechanism is not the same. In this work, we analyze the multi-wavelength interferometric visibilities of dust continuum observations obtained with ALMA and VLA for six transition disks: CQTau, UXTau A, LkCa15, RXJ1615, SR24S, and DMTau, and calculate brightness radial profiles, where diverse emission morphology is revealed at different wavelengths. The multi-wavelength data is used to model the spectral energy distribution and compute constraints on the radial profile of the dust surface density, maximum grain size, and dust temperature in each disk. They are compared with the observational signatures expected from various physical mechanisms responsible for disk cavities.
The observational signatures suggest that the cavities observed in the disks around UXTau A, LkCa15, and RXJ1615 could potentially originate from a dust trap created by a companion. Conversely, in the disks around CQTau, SR24S, DMTau, the origin of the cavity remains unclear, although it is compatible with a pressure bump and grain growth within the cavity.
\end{abstract}


\section{Introduction}
\label{sec:intro}
Protoplanetary disks are dynamic objects. Physical processes such as photo-evaporation \citep[e.g.,][]{Alexander_2006}, accretion \citep[e.g.,][]{Hartmann_2016}, dust migration and segregation \citep[e.g.,][]{Whipple_1972, Weidenschilling_1977,Takeuchi_2002}, dust growth \citep[e.g.,][]{Testi_2014, Birnstiel_2016}, settling \citep[e.g.,][]{Schapler_2004, Avenhaus_2018, Rich_2021}, and disk winds \citep{Konigl_2000, Pascucci_2023}, are taking place in protoplanetary disks. These mechanisms have an important impact on the disk evolution, the spatial distribution of solids and gas, and thus, on the observed disk morphologies at different wavelengths.

In the last decade, high angular resolution observations of protoplanetary disks have unveiled a diversity of large and small spatial scale sub-structures, including vortices \citep[e.g.,][]{Brown_2009, Andrews_2009, Casassus_2013, Perez_2014, Robert_2020, Varga_2021}, cavities \citep[e.g.,][]{Francis_2020}, rings and gaps \citep[e.g.,][]{Huang_2018}, spiral arms \citep[e.g.,][]{Mouillet_2001, Muto_2012, Perez_2016, Paneque_2021}, and circumplanetary disks \citep[][]{Isella_2019b, Benisty_2021, Christiaens_2024}. The origin of the observed morphologies has been attributed to various physical mechanisms such as gravitational instabilities \citep[][]{Toomre_1964}, magneto-rotational instability (MRI, \citealt{Balbus_1991}), dust traps \citep[][]{Whipple_1972, Klahr_1997}, planet-disk interactions \citep[e.g.,][]{Goldreich_1979, Lin_1979}, and snow lines \citep[e.g.,][]{Hayashi_1981}, among others.

The existence of large dust cavities \citep[with sizes from a few au to hundreds of au, ][]{Francis_2020} in the so-called ``transition disks" was first inferred by Near Infrared observations \citep{Strom_1989}, and then by their millimeter spectral energy distribution \citep[SED, ][]{Calvet_2002}, long before high angular resolution data became available. The SED of these objects shows an excess emission at millimeter wavelengths, but no excess emission at mid-infrared wavelengths. This distinct signature was initially interpreted as a deficit of hot material in the inner disk region \citep[e.g.,][]{Strom_1989, Calvet_2002, Brown_2007}, an interpretation that was later confirmed through subsequent high angular resolution observations \citep{Andrews_2011}.

The physical origin of large cavities in transition disks, remains unclear. Nonetheless, various proposed mechanisms aim to replicate these observed dust cavities. As pointed out by \cite{vanderMarel_2023}, these mechanisms include the presence of companions \citep{Artymowicz_1996}, dead zones \citep[][]{Regaly_2012, Flock_2015}, photo-evaporation \citep{Alexander_2006, Garate_2021}, and/or grain growth \citep{Dullemond_2001}. In scenarios involving companions, an already formed planet or a close binary can carve a significant gap in the disk, inhibiting the radial migration of dust grains and resulting in the formation of a large cavity \citep[e.g.,][]{Dong_2015}. This mechanism leads to the creation of a dust trap at the outer edge of the cavity due to the pressure gradient \citep{Pinilla_2012}.

A large cavity can also form by a dead zone, where the particle flow halts due to a jump in gas surface density and a pressure maximum \citep[][]{Flock_2015}. Unlike the mechanism involving a companion, the dead zone scenario does not predict a gas gap within the cavity \citep{Pinilla_2016}.

On the other hand, dust cavities can also form via dust growth. In this mechanism, large grains are expected to have a fast radial migration towards the inner region of the disk \citep{Takeuchi_2002}, resulting in a deficit of dust mass within the disk cavity. Additionally, very large grains (with sizes exceeding about 1 cm) have a low opacity at millimeter wavelengths \citep[e.g.,][]{Draine_2006}, thereby reducing the dust continuum emission within the cavity. However, even in the most optimistic scenario where dust growth is very efficient, this mechanism alone has been shown to be insufficient for creating large cavities within 5 Myrs of dust evolution \citep{Birnstiel_2012b}. Thus, in addition to dust growth, another physical mechanism producing a pressure bump is needed to reproduce the significant drop of millimeter emission within the cavities of transition disks \citep{Birnstiel_2012b}.

Different observational signatures are expected depending on whether the large cavities result from a companion, dead zone, photo-evaporation, or enhanced grain growth. In scenarios involving a dust trap (created by a companion or a dead zone), one would expect an increase in both surface density and maximum grain size at the outer border of the cavity. Conversely, if enhanced grain growth and a pressure bump are responsible for the large cavity, we would expect to infer a prevalence of large dust grains in the inner region of the disk and dust grains being trapped at the border of the cavity.

Observational constraints on the dust surface density, maximum grain size, and temperature can offer valuable insights into the origin of cavities within transition disks. Fortunately, these constraints can be derived from polarized observations \citep{Kataoka_2016b, Kataoka_2016, Yang_2016, Lin_2024, Yang_2024} or from fitting the SED of the dust continuum emission \citep[e.g.,][]{Perez_2012, Carrasco-Gonzalez_2019,  Macias_2021, Sierra_2021, Liu_2024}. It is known that both methodologies offer different grain size estimations (hundreds for microns from polarization and a few millimeter from the SED). Although some works \citep{Tazaki_2019, Lin_2023, Zhang_2023} have demonstrated that millimeter porous grains can help to alleviate the tension, it has been shown that only the opacity of compact dust grains can reproduce the observed spectral index in disk population synthesis \citep{Delussu_2024}.

In this paper, we analyze the dust continuum emission of six known transition disks: CQTau, UXTau A, LkCa15, RXJ1615, SR24S, and DMTau, with cavity sizes of approximately 50 au, 31 au, 76 au, 30 au, 35 au, and 25 au, respectively \citep{Pinilla_2014, Francis_2020}. One of the goals of this work is to compare the multi-wavelength brightness profile for each disk, and infer dust properties than can provide insights into the origin of the cavities.

The observational data is presented in Section \ref{sec:obs}. The analysis and modeling of dust continuum visibilities for each observation are conducted in Section \ref{sec:vismod}, where high angular resolution radial profiles are computed from the data. Section \ref{sec:multiwave} presents the analysis of multi-wavelength radial profiles, revealing constraints on the dust surface density, maximum grain size, and dust temperature within each disk. The discussion of the physical mechanisms influencing each disk is outlined in Section \ref{sec:discussion}, and the general properties are discussed in Section \ref{sec:origin}, followed by Conclusions in Section \ref{sec:conclusions}.

\begin{table}[h]
    \centering
    \caption{Stellar and disk properties}    
    \begin{tabular}{c|ccccc}
    \hline \hline 
    Source & Distance  & $i$ & $\rm PA$ & $L_*$    & $M_*$ \\
    
           & (pc)  & (deg) & (deg) & ($L_{\odot}$) & ($M_{\odot}$)  \\
    \hline
    CQTau  & 149.4 & 36.3 & 46.5  & 10.0 & 1.63 \\
    UXTau  & 140   & 40.9 & 168.7 & 2.5  & 1.4  \\    
    LkCa15  & 157.2 & 49.7 & 62.1  & 1.3  & 1.32 \\
    RXJ1615 & 155.6 & 46.9 & 146.1 & 1.3  & 1.1  \\    
    SR24S   & 114   & 47.8 & 26.6  & 2.5  & 0.87 \\    
    DMTau  & 144.0 & 35.7 & 157.1 & 0.2  & 0.39 \\

    \hline \hline
    \end{tabular}
    \newline
    References: Inclination and geometry are computed in this work. The distances are taken from \cite{Gaia_2020}, and the luminosity and mass of the star from: \cite{Natta_2006, Espaillat_2010, Wahhaj_2010, Andrews_2011, Manara_2014, vanderMarel_2015, Ubeira_2019,   Francis_2020}.
    \label{tab:Stellar}
\end{table}

\section{Observations}\label{sec:obs}
\begin{figure*}
    \centering
    \includegraphics[width=\textwidth]{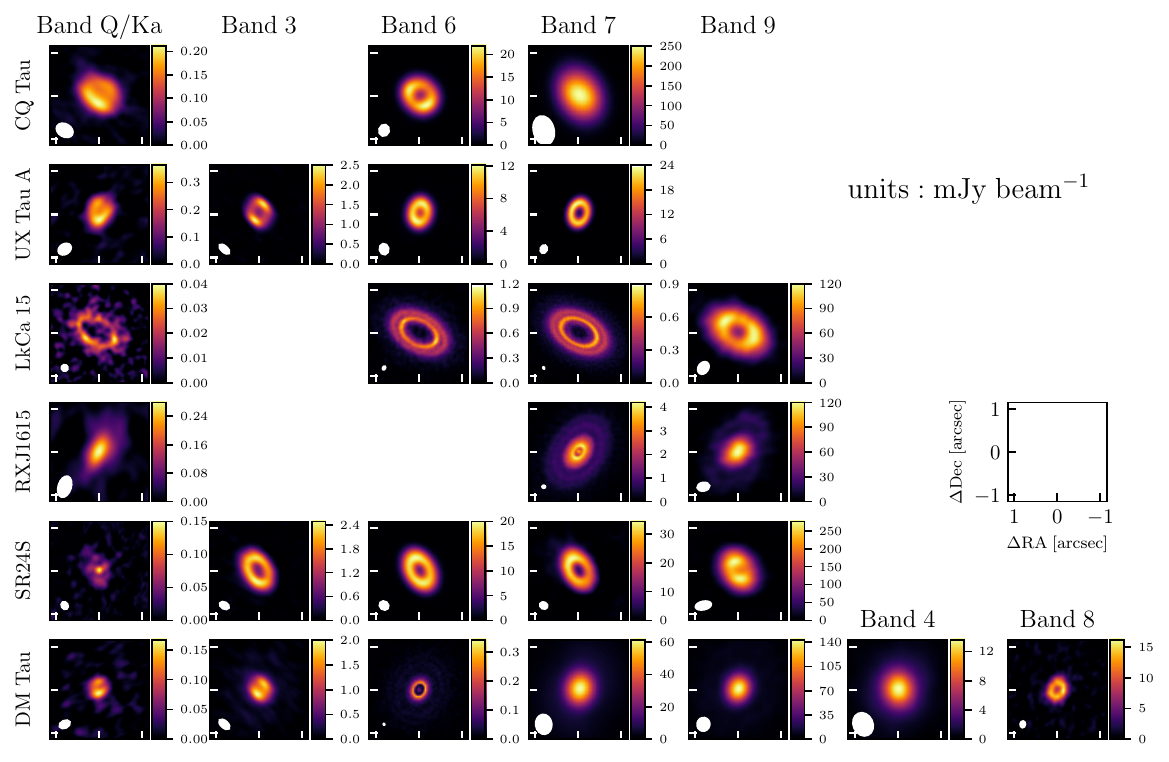}
    \caption{ALMA (from Band 3 to Band 9) and VLA (Band Q or Band Ka) continuum maps of the 6 transition disk in our sample. The beam size is shown in the bottom left corner of each panel. xticks and yticks in all panels are separated by 1 arcsec.}
    \label{fig:Observations}
\end{figure*}

We used 26 archival dust continuum observations of 6 transition disks: CQ Tau, UX Tau, LkCa15, RXJ1615, SR24S, and DMTau. The dust continuum observations spans from optically thick emission (observed in ALMA Band 9 or Band 7) to optically thin emission (observed in VLA Band Q or Band Ka) in all disks, with beam sizes of tens and hundreds of milli-arcseconds.
The distance, stellar parameters, and geometry in each disk are summarized in Table \ref{tab:Stellar}. A summary of the multi-wavelength datasets, final imaging properties, and project codes for both ALMA and VLA observations can be found in Table \ref{tab:Observations}, and a gallery of the dust continuum  images in Figure \ref{fig:Observations}.

The ALMA data were self-calibrated using CASA 6.4.3.27. All spectral windows and scans are combined to improve the signal-to-noise ratio (SNR) using \texttt{combine=`spws, scans'} within the CASA \texttt{gaincal} task. We began applying phase self-calibration with an initial solution time interval equal to the total observation time.  The new self-calibrated data were imaged and the signal-to-noise ratio (SNR) is computed. If SNR improves by at least a factor of 5\%, we continue applying successive phase self-calibration iterations with a smaller solution time interval (it decreases by a factor of 3 in consecutive iterations). Throughout each self-calibration step, we utilized \texttt{applymode=`calonly'} to calibrate data only, and do not apply flags from solutions. A final amplitude self-calibration iteration with an infinitive solution time interval (\texttt{solint=`inf'}) and combining spectral windows and scans (\texttt{combine=`spws,scans'}) is also applied to the data. The calibrated VLA data were taken from the Disks@EVLA Collaboration\footnote{ \href{https://safe.nrao.edu/evla/disks/}{https://safe.nrao.edu/evla/disks/}}. The calibration process for the Disks@EVLA data is provided in Section 2.2 of \cite{Perez_2015} or Section 2 of \cite{Andrews_2014}.
For imaging, we employed the \textsc{tclean} task with a multi-scale CLEAN algorithm with scales corresponding to point sources, 1, 2, and 3 times the beam Full Width at Half Maximum (FWHM). Additionally, we applied a mask covering the disk emission.

\begin{table*}
    \centering
    \caption{Properties of the millimeter continuum images.}    
    \begin{tabular}{c|cccccccccc}
    \hline \hline 
    Source & ALMA/  & $\nu$ & Baselines& On-Source & robust & Synthesized Beam & rms Noise & ALMA/VLA    \\
           & VLA Band  & (GHz) &  (m-km) & Time (min) & & (mas $\times$ mas; deg) & ($\mu \rm Jy \ beam^{-1}$) & Project Code \\
    
    \hline
    CQTau & 7 & 338 &  21 - 0.4 & 22.1 & 0.5 & 674 $\times$ 469; 15 & 530.0 & 2011.0.00320.S \\
           & 6 & 225 & 3 - 8.5 & 93.2 & 0.5 &267 $\times$ 231; -16& 35.0 & 2013.1.00498.S \\
           & Ka & 34 & 80 - 11.1 & 115.1 & 2.0 & 401 $\times$ 293; 61 & 7.7 & AC982	\\
           
    \hline
    UXTau A & 7  & 344 &  16 - 1.5  & 54.4 & 0.5 &193 $\times$ 151; -22 & 49.0 & 2015.1.00888.S\\
            & 6 & 232   & 15 - 1.6  & 14.1 & 0.5 &253 $\times$ 209; 11  & 37.3 & 2013.1.00498.S \\
            & 3 & 96    & 18 - 7.6  &  8.6 & 0.5 & 296 $\times$ 140; 49  & 27.0 & 2016.1.01042.S \\
            & Q & 42    & 78 - 11.1 &  144.2& 2.0 & 315 $\times$ 235; -57 & 14.9 & AC982	\\
           
    \hline
    LkCa15 & 9 & 689 &  21 - 0.4 & 17.1 & 0.5 & 306 $\times$ 229; -35 & 1650.0 & 2011.0.00724.S\\
            & 7 & 340 & 34 - 8.5  & 162.1 & 0.5 & 47 $\times$ 27; 23   &  16.5& 2018.1.00350.S\\
            & 6 & 229 & 15 - 12.7 & 140.2 & 0.5 & 84  $\times$ 57; -30  & 14.4 & 2018.1.01255.S\\
            & Q & 44  & 78 - 27.9 & 1129.0 & 2.0 & 149 $\times$ 144; 64  & 4.2  & AC982,13B-381,12B-196 \\          
    \hline
    RXJ1615 & 9 & 689 &  21 - 0.4 & 22.2 & 0.5 & 288 $\times$ 193; -80 & 1750.0 & 2011.0.00724.S \\
            & 7 & 339 &  18 - 3.7  & 59.4 & 0.5 & 73 $\times$ 61; 84 & 28.0 & 2012.1.00870.S \\
            & Q & 44 & 243 - 11.1 & 152.4 & 2.0 & 498 $\times$ 294; -17 & 16.2 & 13B-381	\\
    
    \hline
    SR24S & 9 & 689 &  21 - 0.4  & 25.2 & 0.5 & 368 $\times$ 190; -78 & 1910.0 & 2011.0.00724.S\\
          & 7 & 337 &  15 - 1.6  & 2.0  & 0.5 & 183 $\times$ 155; 61 & 367.1   & 2013.1.00157.S\\
          & 6 & 231 &  15 - 1.6  & 21.7 & 0.5 & 222 $\times$ 192; 51 & 67.1  & 2013.1.00091.S\\
          & 3 & 109 &  19 - 9.5  & 7.2 & 0.5 & 233 $\times$ 155; 61 & 33.4  & 2016.1.01042.S\\
          & Ka& 34  & 78 - 36.6 & 327.5 & 2.0 & 180 $\times$ 148; 31 & 7.7 & AC982	\\
          
    \hline
    DMTau & 9 & 676 &  15 - 0.5  & 79.9 & 0.5 & 309 $\times$ 287; -11 & 674 & 2016.1.00565.S\\
        & 8 & 480 &  15 - 1.1  & 11.1 & 0.5 & 150 $\times$ 115; -6 & 660 & 2015.1.01137.S\\
        & 7 & 318 &  15 - 0.9  & 79.5 & 0.5 & 455 $\times$ 366; 7  & 139 & 2016.1.00565.S\\
        & 6 & 232 &  15 - 15.2  & 134.3 & 0.5 & 32  $\times$  21; 33 & 8.4 & 2018.1.01755.S \\
        & 4 & 145 &  15 - 1.1  & 2.0 & 0.5 & 546 $\times$ 448; 24 & 84 & 2015.1.00296.S\\
        & 3 & 109 &  19 - 7.6 & 8.6 & 0.5 & 296 $\times$ 158;  48 & 27  & 2016.1.01042.S\\
        & Q & 42 & 78 - 22814 & 378.3 & 2.0 & 259 $\times$ 173; -65 & 9.0 & AC982\\
          
    \hline \hline
    \end{tabular}
    \label{tab:Observations}
\end{table*}

\section{Visibility modeling} \label{sec:vismod}
We fit the de-projected continuum visibilities of the dust continuum emission to extract the highest angular resolution from the observational data. The de-projection of the visibilities is done using the \textsc{Frankenstein} tools\footnote{available at \href{https://github.com/discsim/frank}{https://github.com/discsim/frank}} in \cite{Jennings_2020},  where the geometric projecting effects due to the orientation of the interferometer's baselines are corrected. The de-projection consists of rotating the baseline coordinates using the disk's position angle, and then stretching the baseline coordinates along the minor axis of the disk using the cosine of the disk inclination.

We assume azimuthal emission symmetry, which is a first order approximation of the disk morphologies in this work, as shown in Figure \ref{fig:Observations}, although it is known that some these disk present some degree of azimuthal asymmetry \citep[e.g., DMTau band Q observations in][]{Hashimoto_2021}.

The interferometric visibilities (which for a given uv-distance, corresponds to the Real and Imaginary part of the visibility and its associated weight) are extracted from the self-calibrated measurement sets using a modified version from the \textit{export\_uvtable} function in \cite{uvplot}. In our version, the uv-distances are normalized using the wavelength of each channel and spectral window, instead of the average wavelength of the whole data set.
The visibility modeling is done using a non-parametric fit (Section \ref{sec:non-parametric}) for observations with high signal-to-noise ratio (SNR), and with a parametric model for low SNR observations (Section \ref{sec:parametric}). In both cases, constraints to the phase shift and geometry of each observation (Section \ref{sec:geometry}) are necessary when de-projecting and fitting the visibilities.

\subsection{Non-parametric fit}
\label{sec:non-parametric}
The non-parametric visibility modeling is performed using \textsc{Frankenstein} \citep{Jennings_2020}, where the disk brightness (represented as a Fourier-Bessel series) is reconstructed by fitting the real part of the visibilities as a function of the uv-distance (baselines). The imaginary part of the visibilities, that mainly traces the emission from non-symmetric structures, is not used by \textsc{Frankenstein} given the azimuthal symmetry assumption (the imaginary part of the visibilities in our observations tends to zero when averaged over all orientations).

The reconstructed radial profile depends on two hyper-parameters: $\alpha$, $w_{\rm smooth}$. The former mimics the signal-to-noise (SNR) threshold below which the data are not used when fitting the visibilities, and the latter controls the smoothness of the fit to the power-spectrum \citep[see full details about the visibility fit in][]{Jennings_2020, Jennings_2021}.

These hyper-parameters do not have strong effects on the reconstructed brightness profiles for disks with a high SNR; however, this is not the case for low SNR data, where a particular combination of $\alpha$, $w_{\rm smooth}$ can imprint oscillatory artifacts on the reconstructed brightness profile. For this reason, we fit all the observations using a grid of hyper-parameters varying between $\alpha = [1.05-1.30]$, and $ \log_{10}(w_{\rm smooth}) = [-4, -1]$, and we choose by visual inspection the pair of hyper-parameters that minimize the oscillatory artifacts. 

The longest wavelengths in our sample (Band Q and Ka from VLA) have a low SNR at all uv-distances. Thus, for these long wavelength observations, we use \textsc{Frankenstein} on the binned visibilities (binsize of 10 k$\lambda$).

The chosen hyper-parameters for each data set are summarized in Table \ref{tab:Visibility}. Additional fit parameters (e.g., maximum fitting radius, number of radial points to sample the radial profile) have no important effects on the reconstructed brightness profile if the recommendations described in \cite{Jennings_2020} are taken into account. We fit the data in the logarithmic brightness space, which helps to reduce the oscillatory artifacts in the reconstructed brightness profile, and prevents negative values in the reconstructed brightness profile. However, we found no important differences when fitting the data in a linear brightness space.

\subsection{Parametric fit}
\label{sec:parametric}
The \textsc{Frankenstein} models for VLA data were computed by binning the visibilities. Since this procedure results in information loss, we also directly fit the non-binned VLA visibilities using a parametric model with a given intensity profile $I(r)$. The visibilities of the parametric model are computed using a Hankel transform $\mathcal{H}$  \citep{Pearson_1999}:

\begin{equation}
    V(q) = \mathcal{H} (I(r)) = 2\pi \int _{0}^{\infty}I(r) J_0(2\pi q r) dr, 
\end{equation}
where $r$ is the disk radius, $J_0$ is the zeroth-order Bessel function of the first kind, $q=\sqrt{u^2+v^2}/\lambda$ is the de-projected baseline, $\lambda$ is the wavelength, and $u,v$ are the coordinates of the baseline vectors.

For each disk, the morphology of the parametric models is motivated by the brightness distribution inferred at ALMA wavelengths with \textsc{Frankenstein} (an extended discussion on the disk morphology is done for each particular disk in Section \ref{sec:discussion}). We also consider other physical processes that can contribute to the emission at VLA wavelengths, as free-free emission from ionized gas. However, we assume that the emission from spinning nanometer-size dust \citep{Hoang_2018} does not contribute to the observed emission, as scattering properties (taken into account in this work) can mimic similar effects to those expected from spinning dust \citep{Sierra_2020}, making it difficult to differentiate between the two processes.

The free-free emission is modeled by including a compact emission point source at the disk center, simulating the expected compact free-free emission at these long wavelengths reported by previous centimeter observations of UXTau A, LkCa15, RXJ1615, and SR24S in \cite{Zapata_2017}. For DM Tau, no compact free-free emission was detected in \cite{Zapata_2017}, however \cite{Terada_2023} found that the free-free emission from ionized gas in DM Tau is variable. We include the free-free component in all disks, and then we compare its inferred flux with previous estimations in Appendix \ref{app:parametric_models}.

The parameter space of the free parameters describing the parametric fit is explored using a Markov chain Monte Carlo \citep[MCMC, implemented in the Python library \textsc{emcee},][]{Foreman_2013}. The best fit parametric model is found by maximizing the logarithm of the likelihood function, given by

\begin{eqnarray}
\nonumber \ln \mathcal{L} &=& -\frac{1}{2} \sum_j W_j \left[\rm   ( ReV^{obs}_j - ReV^{mod}_j)^2 + \right. \\
 & &  \left. (\rm ImV^{obs}_j - ImV^{mod}_j)^2  \right],
\label{eq:posterior_vis}
\end{eqnarray}
where $W_j$ are the weights of the visibilities, and $\rm V^{obs}_j, V^{model}_j$ are the observed and model visibilities, respectively.

\subsection{Phase shift and Geometry}
\label{sec:geometry}
The offset of the disk center with respect to the phase center ($\rm \delta Ra, \delta Dec$), disk inclination ($i$), and disk position angle (PA) are fitted for each observation.

In most of the cases, the offset is computed by minimizing the imaginary part of the visibilities \citep[e.g.,][]{Isella_2019}. However, in some cases (e.g. SR24S with some small azimuthal asymmetries), the obtained phase shift does not match with the disk center (visual inspection), thus, in a few cases, we fit a 2D Gaussian function (in the image plane) using a very low angular resolution image (using the \textsc{uvtaper} task in CASA), where the azimuthal asymmetry vanishes. The methodology used for each observation is summarized in Table \ref{tab:Visibility}.

The disk inclination and position angle are fitted by minimizing the spread of the de-projected visibilities for circular annuli in the visibility space \citep[e.g.,][]{Isella_2019}. Error bars are computed from the percentiles 16th and 84th.
The inferred disk inclination for a certain disk is consistent within the error bars. However, variations can also occur because disks are not perfectly settled to the midplane, depending on factors such as grain size and turbulent state \citep[e.g.,][]{Dubrulle_1995, Schapler_2004}. Additionally, the contribution of the upper layers becomes more significant at smaller wavelengths \citep[e.g.,][]{Sierra_2019}.

The inferred disk geometry for all the ALMA observations are summarized in Table \ref{tab:Visibility}. The geometry derived from the VLA data is poorly constrained and is not included in this Table.

One of the goals in this paper is to compare the brightness radial profiles using multi-wavelength observations of each disk. For this reason, the geometry of each disk is fixed for all wavelengths.
The chosen disk geometry is based on the following criteria: 1- The contribution from the disk mid-plane should be important (long wavelengths where the emission is optically thin are good candidates). 2- The angular resolution (or uv-distances) should be good enough to resolve the disk geometry. 3- High SNR (i.e., the disk geometry is well constrained). The latter discards the Band Q and Ka observations, where the geometry is poorly constrained. Thus, our final decision on the disk geometry is based on Band 6, 3, 6, 7, 6, 6, for CQTau, UXTau A, LkCa15, RXJ1615, SR24S and DMTau, respectively.

\begin{table*}
\centering
\caption{Visibility modeling parameters.}    
\begin{tabular}{c|ccccccccc}
\hline \hline 
Source & Band & $\alpha$ & $\log_{10}(w_{\rm s})$ & $i^{(*)}$ & PA$^{(*)}$ & ChanWidth &  Centering & Resolution  & Multi-wave \\
 &  & & & (deg) & (deg) & (MHz) &  methodology & (mas) & analysis?\\    

\hline
CQTau & 7 & 1.3 & -1 & $37.9_{-0.3}^{+0.5}$ & $46.7_{-0.1}^{+1.8}$ & 29 &  Min vis scatter &  255.3 & No \\
       & 6 & 1.3 & -4 & $36.3^{+0.1}_{-0.6}$  & $46.5^{+0.9}_{-0.4}$ & 500&  Gaussian fit & 149.6  & Yes \\
       & Ka & 1.05 & -4 & - & - & 128  & Gaussian fit &  238.4 & Yes\\
       
\hline
UXTau A & 7 &  1.05 & -1 & $41.9_{-0.1}^{+0.5}$ & $168.0_{-0.5}^{+0.1}$ & 500 & Min vis scatter &  75.1 & Yes \\
        & 6 & 1.3  & -1 & $41.9_{-0.7}^{+0.1}$ & $166.6_{-0.2}^{+0.8}$ & 117 & Min vis scatter & 136.5 & Yes \\
        & 3 & 1.05 & -4 & $40.9_{-0.1}^{+0.6}$ & $168.7_{-0.5}^{+1.0}$ & 500 & Min vis scatter & 112.3 & Yes \\
        & Q & 1.05 & -4 & - & - & 128 & Gaussian fit &  157.0 & Yes\\
               
\hline
LkCa15 & 9 & 1.2 & -4 & $50.8_{-0.8}^{+0.2}$ & $61.7_{-0.1}^{+2.0}$ & 375 & Min vis scatter & 171.9 & No \\
        & 7 & 1.3  & -1 & $50.5_{-0.5}^{+0.1}$ & $61.6_{-0.2}^{+0.1}$ & 250 & Min vis scatter & 17.7 & Yes \\
        & 6 & 1.3  & -1 & $49.7_{-0.2}^{+0.1}$ & $62.1_{-0.1}^{+0.4}$ & 250 & Min vis scatter &  21.5 & Yes\\
        & Q & 1.05 & -4 & - & - & 128 & Gaussian fit &  120.0 & Yes \\
\hline
RXJ1615 & 9 & 1.05 & -4 & $40.6_{-0.9}^{+9.2}$ & $150.0_{-5.0}^{+0.1}$ & 375 & Min vis scatter &  191.4 & No \\
        & 7 & 1.3 & -1 & $46.9_{-0.9}^{+0.1}$ & $146.1_{-0.1}^{+0.6}$ & 235 & Min vis scatter &  34.2 & Yes\\
        & Q & 1.05 & -4 & - & - & 128 & Gaussian fit &  167.0 & Yes \\
\hline
SR24S & 9 & 1.3 & -1 & $48.7_{-0.6}^{+0.2}$ & $27.2_{-0.5}^{+0.1}$ & 375 & Gaussian fit & 157.2 & No \\
      & 7 & 1.3 & -1 & $48.6_{-0.2}^{+0.1}$ & $27.3_{-0.3}^{+0.1}$ & 16 & Gaussian fit & 118.2 & Yes \\
      & 6 & 1.3 & -1 & $47.8_{-0.1}^{+0.1}$ & $26.6_{-0.6}^{+0.1}$ & 469 & Gaussian fit & 117.2 & Yes \\
      & 3 & 1.3 & -1 & $44.8_{-0.1}^{+2.4}$ & $25.1_{-0.3}^{+4.0}$ & 63 & Gaussian fit & 98.6 & Yes\\
      & Ka & 1.05 & -4 & - & - & 128 & Gaussian fit &  120.1 & Yes \\
      
\hline
DMTau & 9 & 1.05 & -4 & $36.6_{-0.3}^{+0.1}$ & $156.3_{-1.1}^{+0.1}$ & 234 & Min vis scatter & 204.0 & No\\
    & 8 & 1.05 & -1 & $39.9_{-9.3}^{+0.4}$ & $153.3_{-7.3}^{+4.9}$ & 125 & Min vis scatter & 111.9 & Yes \\
    & 7 & 1.3 & -1 & $40.1_{-4.7}^{+1.8}$  & $153.1_{-4.2}^{+0.9}$ & 63 & Min vis scatter & 153.8 & No \\
    & 6 & 1.3 & -1 & $35.7_{-0.1}^{+0.1}$  & $157.1_{-0.1}^{+0.4}$ & 469 & Min vis scatter & 14.2  & Yes \\
    & 4 & 1.1 & -1 & $32.7_{-0.6}^{+0.1}$   & $156.0_{-6.9}^{+0.2}$ & 63 & Min vis scatter & 273.4 & No \\
    & 3 & 1.3 & -1 & $34.8_{-0.6}^{+2.6}$   & $155.2_{-4.9}^{+0.5}$ & 63 & Min vis scatter & 72.8  & Yes \\
    & Q & 1.1 & -4 & - & - & 128 & Gaussian fit & 69.6 & Yes \\
\hline \hline
\end{tabular}
\\
$^{(*)}$ The inferred inclination ($i$) and position angle (PA) for each observation are reported in this table. However, both values are fixed to those summarized in Table \ref{tab:Stellar} during the visibility modeling.
\label{tab:Visibility}
\end{table*}

\subsection{Results from the visibility modeling} \label{sec:Results_Visibilities}
\begin{figure*}
    \centering
    \includegraphics[width=0.9\textwidth]{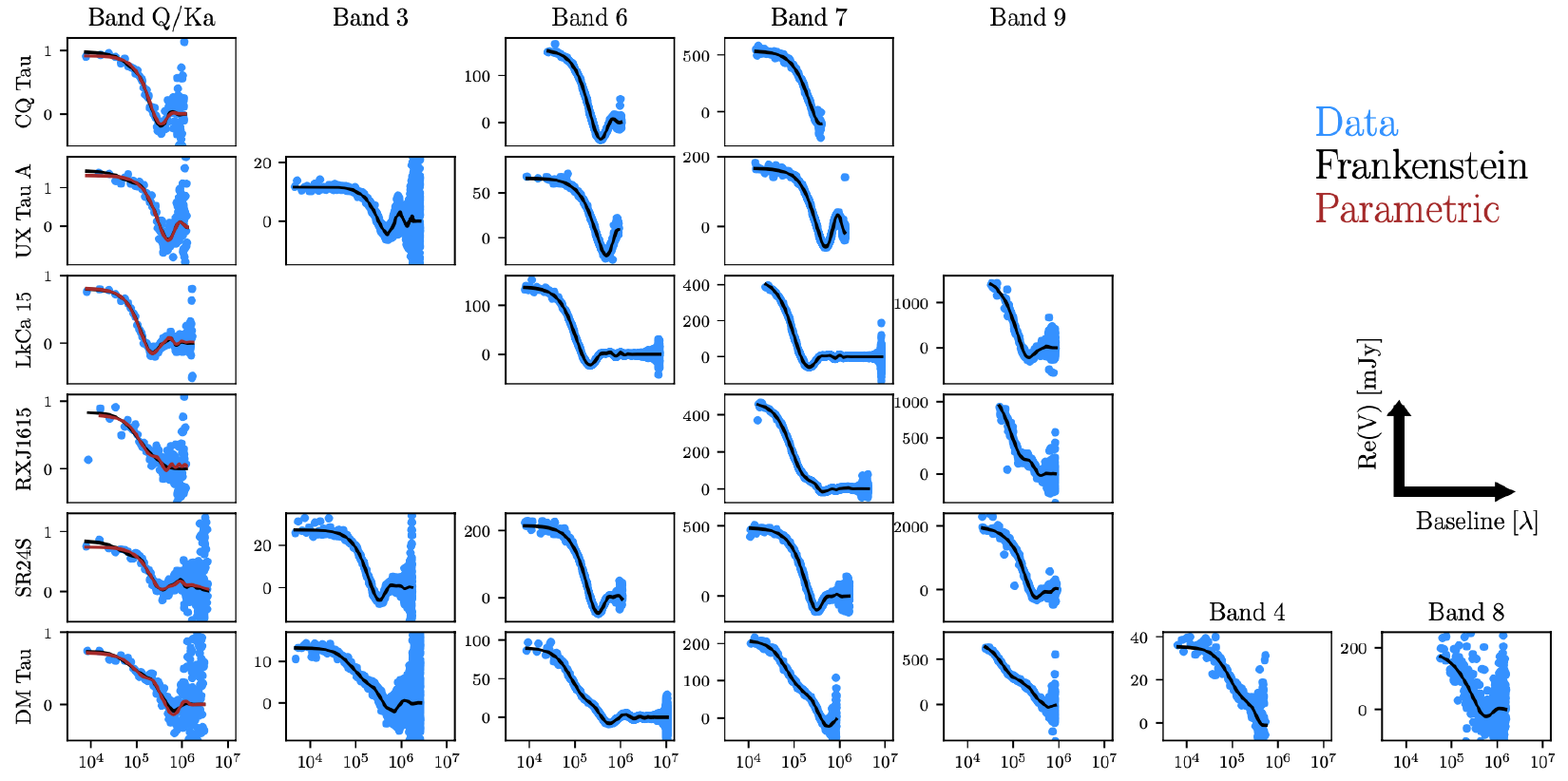}
    \caption{Real part of the dust continuum visibilities (blue points) as a function of baseline for each observation in this work. The black/red solid line is the \textsc{Frankenstein}/Parametric visibility model.}
    \label{fig:Visibilities}
\end{figure*}

Figure \ref{fig:Visibilities} shows the visibility modeling results for the observations in our sample. In all cases, the axi-symmetric visibility model (from Frankenstein or parametric) is a good representation of the observed visibilities. Note that the uv-coverage of each observation is diverse. For example, for LkCa15, the maximum baseline for Band 6 and 7 is close to 10M$\lambda$, but for Band 9 it is close to 1M$\lambda$. This limits the disk sub-structures that can be inferred from observations.

\begin{figure*}
    \centering
    \includegraphics[width=0.9\textwidth]{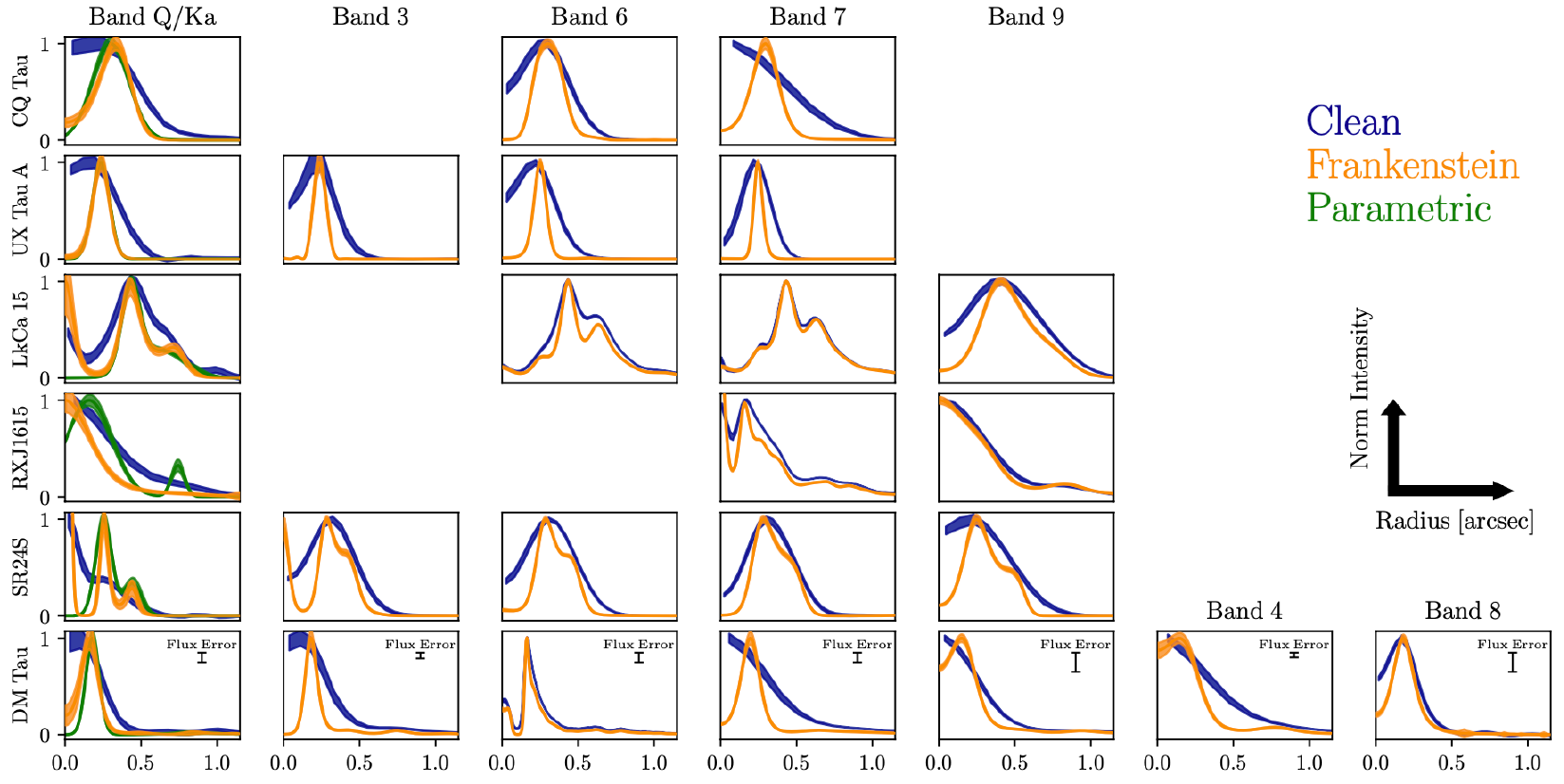}
    \caption{Normalized brightness radial profile for each observation in our sample. The blue lines are the azimuthal average computed from the image plane in Figure \ref{fig:Observations}, the orange lines are the non-parametric radial profile computed with \textsc{Frankenstein}, and the green lines for the VLA observations are the parametric radial profiles. The vertical error bars in the bottom panels show the uncertainty from the flux calibration at each Band.}
    \label{fig:Vis_RadProfile}
\end{figure*}

Figure \ref{fig:Vis_RadProfile} shows the normalized brightness radial profile for each observation from the visibility fit and from the image plane. The radial profiles in the image plane were computed using \textsc{GoFish} \citep{GoFish} and the same offset and disk geometries than in the visibility fits. We assume a filling factor \citep{Thompson_2017} of 1 everywhere. This assumption is valid at large radius, where the solid angle of concentric rings is larger than that of the beam size. However, in the most inner region ($\lesssim$ 50 mas), the radial profiles from the image plane would need a filling factor correction. Nevertheless, the radial profiles from the image plane are only used to visualize the differences compared with the radial profiles computed from the visibility fit, and they are not used in the rest of the manuscript.

The error bars $(\Delta_{\rm img})$ in the radial profile from the image plane $(I_{\rm img})$ do take into account the azimuthal intensity asymmetries, while they are not originally taken into account in the error bars from the visibility fits. The ratio of azimuthal asymmetry $(\Delta_{\rm img}/I_{\rm img})$ is small in all cases, confirming that the axi-symmetric assumption is a good approximation of the disk morphologies.
We add the azimuthal asymmetries error bars computed in the image plane to the radial profile uncertainty computed from the visibility fit ($\Delta_{\rm vis}$) as: $\Delta I = \sqrt{(\Delta_{\rm vis})^2 + (I_{\rm vis} \Delta_{\rm img}/I_{\rm img})^2}$, where $I_{\rm vis}$ is the intensity radial profile from the visibility fit. These uncertainties are included in the radial profiles of Figure \ref{fig:Radial_Profiles}.
The bottom panels of the same Figure show the flux calibration error bar for all wavelengths. This uncertainty dominates by far the uncertainty in each radial profile.

The radial profiles from the visibility fit reveal higher contrasts between rings and gaps compared to the azimuthal radial profiles computed from the \textsc{clean} images (Figure \ref{fig:Observations}). For example, in the brightness radial profile of LkCa15 in Bands 6 and 7, we clearly identify three main rings from the visibility fit (at $0.277^{\prime\prime}, 0.433^{\prime\prime}, 0.635^{\prime\prime}$, as also reported by \cite{Long_2022}). Additionally, there is a tentative faint ring around $\sim 1.07^{\prime \prime}$, which is also visible in the radial profile of the Band Q image, albeit at a radius of $\sim 0.99^{\prime \prime}$. Note that the width of the rings at $0.433^{\prime\prime}, 0.635^{\prime\prime}$ in Band 6 looks narrower than those at Band 7, possibly due to dust traps \citep{Pinilla_2012}. Additionally, the contrast between the rings and gaps is more pronounced in Band 6, likely due to optical depth effects. These characteristics and differences are crucial when constraining the radial profiles of dust properties in Section \ref{sec:multiwave}.

The parametric model for all disks was chosen based on the morphology observed at ALMA. For example, in LkCa15, the parametric model is given by two Gaussian functions at a fixed radial position of $0.433^{\prime\prime}, 0.635^{\prime\prime}$, and a point source at the disk center. We also tested the inclusion of the ring structures at $0.277^{\prime\prime}$ and $0.99^{\prime\prime}$. However, the chi-square value ($\chi^2$) does not exhibit a significant improvement when these rings are added to the model.

The radial profiles in Figure \ref{fig:Vis_RadProfile} are shown at their intrinsic angular resolution. Starting from Section \ref{sec:multiwave} and beyond, the radial profiles for a specific disk are convolved to the same angular resolution, corresponding to the lowest angular resolution among the multi-wavelength observations of each disk.

As mentioned in Section \ref{sec:parametric}, the \textsc{Frankenstein} visibility model is influenced by bin size and hyper-parameters for the VLA observations with low SNR. Conversely, the parametric fit provides a robust radial profile. Consequently, the parametric model is selected as the radial profile model for VLA data across all disks. Table \ref{tab:parametric_fit} in Appendix \ref{app:parametric_models} provides a summary of the chosen parametric models for each disk and the best fit parameters.

Figure \ref{fig:models} in Appendix \ref{app:best_models} shows the \textsc{CLEAN} image of the disk models (sampled at the same uv-coverage than the observations). The color scale is the same than each panel in Figure \ref{fig:Observations}. The difference between observations and model (Residuals) is shown in Figure \ref{fig:residuals}. Although most of them have low non-axi-symmetric residuals, some exceed $+5\sigma_{\rm rms}$ or fall below $-5\sigma_{\rm rms}$. Various factors may contribute to these residuals, such as the disk potentially being brighter at the inner wall of the ring on the far side due to disk inclination, uncertainties in parameters like disk inclination, position angle, disk center, or elevated emission surfaces  \citep{Andrews_2021}. Also note that most of the azimuthal asymmetries observed at very long wavelengths (VLA observations) have a SNR $<3$, making it difficult to distinguish them from thermal noise.
The specific disk structures are further discussed in Section \ref{sec:discussion}.
\vspace{1cm}

\section{Constraints on Dust Properties} \label{sec:multiwave}

The spatially-integrated dust continuum emission at millimeter wavelengths has been used to infer dust properties of protoplanetary disks, including mean temperature, optical depths, average maximum grain size, and dust mass. This is typically achieved by fitting a black-body model to their spectral energy distribution \citep[e.g.,][]{Rilinger_2023}.

\begin{figure*}
    \centering
    \includegraphics[scale=0.9]{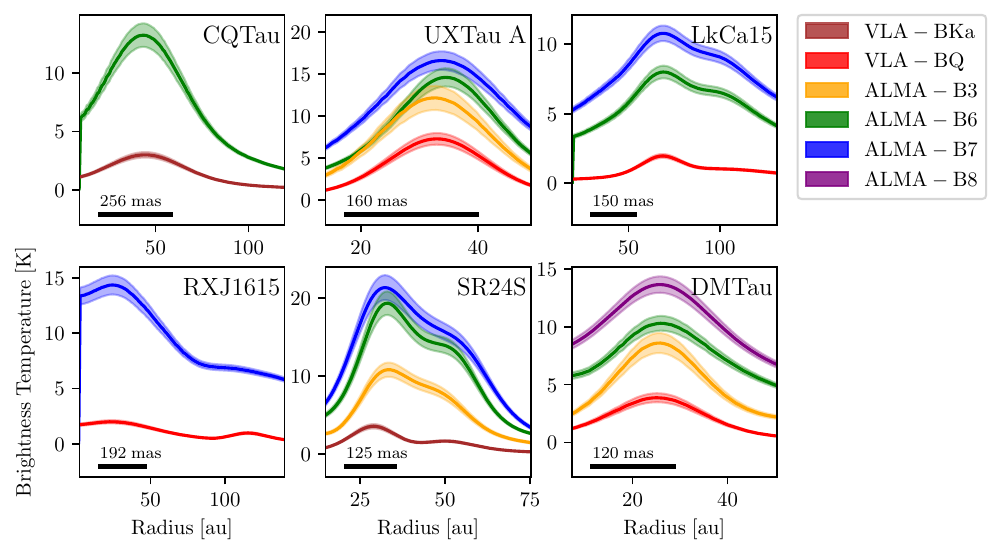}
    \caption{Brightness temperature radial profile for the six transition disks used in the multi-wavelength analysis. The common angular resolution for all wavelengths in each disk is shown as a horizontal black line in each panel. The error bar for each profile is computed from errors from the flux calibration, visibility fit, and azimuthal variations}.
    \label{fig:Radial_Profiles}
\end{figure*}

Owing to the lack of multi-wavelength observations at high angular resolution, the spatial modeling of spectral energy distributions (SED) have only been spatially modeled in a small fraction of disks \citep[e.g.][]{Perez_2012, Liu_2017, Carrasco-Gonzalez_2019, Macias_2021, Sierra_2021}. In this work, we are able to resolve the expected optically thick rings around six transition disks and model the SED as function of radius. Figure \ref{fig:Radial_Profiles} shows the brightness temperature for the six transition disks in this work. The radial profiles of each disk are convolved to the same angular resolution (indicated by the horizontal bar in each panel), determined by the lowest angular resolution detailed in the last column of Table \ref{tab:Visibility}. The effect of radial beam smearing is discussed in Appendix \ref{app:smearing}.

Note that not all wavelengths were included in our multi-wavelength analysis. We opted to exclude observations with low angular resolution (compared with the rest of the multi-wavelength data) and avoided convolving most of the observations using a large beam.
Additionally, we also decided to neglect the short wavelength observations at Band 9 because they likely primarily capture dust emission from the upper layers of the disk, where smaller grain sizes are expected compared to those traced by longer wavelengths in the mid-plane \cite[e.g.,][]{Sierra_2020}. The last column in Table \ref{tab:Visibility} summarizes which datasets are used in the multi-wavelength analysis.

The SED is fitted at each radius using the emergent intensity of the dust continuum emission at frequency $\nu$ in \cite{Sierra_2020}, which is given by
\begin{equation}
    I_{\nu} = B_{\nu}(T_d) \left[1- \exp(-\tau_{\nu}/\mu) + \omega_{\nu} F(\tau_\nu, \omega_\nu) \right],
    \label{eq:Intensity_scatI}
\end{equation}
where $B_{\nu}(T_d)$ is the Planck function evaluated at the dust temperature $T_d$, $\tau_{\nu}$ is the total optical depth (absorption and scattering), $\mu$ is the cosine of the disk inclination, $\omega_\nu$ is the dust albedo, and $F(\tau_\nu, \omega_\nu)$ is the contribution of scattering to the emergent intensity, given by

\begin{eqnarray}
    \nonumber F(\tau_{\nu}, \omega_{\nu}) = \frac{1}{\exp(-\sqrt{3}\epsilon_{\nu} \tau_{\nu} ) (\epsilon_{\nu} -1) - (\epsilon_{\nu}+1)} \times\\
    \nonumber \left[ \frac{1 - \exp(-(\sqrt{3}\epsilon_{\nu}+1/\mu)\tau_{\nu})}{\sqrt{3}\epsilon_{\nu}\mu +1} \right.+ \\
    \left. \frac{\exp(-\tau_{\nu}/\mu) - \exp(-\sqrt{3}\epsilon_{\nu} \tau_{\nu})}{\sqrt{3}\epsilon_{\nu} \mu -1} \right],
    \label{eq:Intensity_scatII}
\end{eqnarray}
where $\epsilon_\nu = \sqrt{1-\omega_\nu}$.
The main effect of scattering (Eq. \ref{eq:Intensity_scatII}) is decreasing the emergent intensity in the optically thick regime, making it possible to explain spectral indices below of 2, as discussed in previous works (e.g., \citealt{Liu_2019, Zhu_2019, Sierra_2020}).

The albedo and absorption dust opacity coefficient $\kappa_{\nu}$ are taken from \cite{Birnstiel_2018}, and the dust grains are assumed to follow a particle size distribution (number of grain sizes per size bin) given by $n(a)da \propto a^{-p}da$, where the minimum grain size is fixed to 0.05 $\mu m$, and the maximum grain size $a_{\rm max}$ is a free parameter. The slope of the particle size distribution $p$ is fixed to 3.0. The expected value from simulations is between 2.5 and 3.5, when the maximum grain size is regulated by drift and fragmentation, respectively \citep{Birnstiel_2012}. However, we tested $p$ values within this range, and the qualitative conclusions are not sensitive to this particular choice.

Equation \ref{eq:Intensity_scatI} can be evaluated given the dust temperature $T_{\rm d}$, maximum grain size $a_{\max}$ (which controls $\omega_{\nu}$ and $\kappa_{\nu}$), and dust surface density $\Sigma_{\rm d}$ (which controls the total optical depth: $\tau_{\nu} = \kappa_{\nu} \Sigma_{\rm d}/(1-\omega_{\nu})$). We use a Markov chain Monte Carlo (MCMC) method \citep[implemented in the \textsc{Python} library \textsc{emcee},][]{Foreman_2013} using 24 walkers and 10000 steps to explore the space parameter of $a_{\rm max}, \Sigma_{\rm d}$, and $T_{\rm d}$ at each radius. The chi-squared in the posterior distribution is defined as
\begin{equation}
    \chi^2 = \sum _n \left( \frac{I_{\nu_n} - I_{\nu_n}^{\rm model}}{\Delta I_{\nu_n}} \right)^2,
\end{equation}
$I_{\nu_n}^{\rm model}$ is the model intensity evaluated from Equation \ref{eq:Intensity_scatI}, and  $\Delta I_{\nu_n}$ is the uncertainty of the n-th intensity. The absolute flux uncertainty is taken into account in $\delta I_{\nu_n}$ at each band according to the ALMA proposer's guide: 5\% in Band 3 and 4, 10\% in Band 6 and 7, and 20\% in Bands 8 and 9. 

The prior is given by $p(a_{\rm max}) \times p(\Sigma_{\rm d}) \times p(T_{\rm d})$, where $p(a_{\rm max})$, $p(\Sigma_{\rm d})$ are the maximum grain size and dust surface density priors, given by a log-uniform distribution within the explored parameter space, and the temperature prior is defined as
\begin{equation}
    p(T_{\rm d}) = \exp \left[ -\frac{1}{2} \left(\frac{T_{\rm prior} - T_{\rm d}}{\sigma_{T_{\rm prior}}}\right )^2 \right],
\end{equation}
i.e., the probability of a model with a temperature $T_{\rm d}$ decreases if it deviates from $T_{\rm prior}$. The latter is computed from the temperature of an passively irradiated disk $T_{\rm prior} = (0.02 L_* / 8 \pi \sigma_{\rm B} r^2)^{1/4}$. We use a wide temperature width $\sigma_{T_{\rm prior}} = 50$ K  to ensure that this prior does not dominate the posterior probability distribution.

The inclusion of this prior decreases the degeneracy between the free parameters, especially when the number of wavelengths used to fit the SED is less than or equal to the number of free parameters.

The explored parameter space is log-spaced between 300 $\mu$m $\leq a_{\rm max} \leq 30$ cm, $10^{-6}$ g cm$^{-2}$ $\leq \Sigma_{\rm d}$ $\leq 10$ g cm$^{-2}$, and $3$ K $\leq T_{\rm d}$ $\leq 300$ K. The smallest value for the maximum grain size is chosen to avoid degeneracy between a small-grain solution ($a_{\rm max} < 300 \mu$m) and large grains ($a_{\rm max} \geq 300 \mu$m). As shown in Figure 4 of \cite{Birnstiel_2018}, the spectral index around 300 micrometers is degenerated for large or small grain sizes. However, we also did some tests including small maximum grain sizes $a_{\rm max} = 10 \ \mu$m, but found that the walkers tend to the large maximum grain size regime after sufficient steps. The inclusion of small maximum grain sizes ($a_{\rm max} < 300 \mu$m) only delays the walkers from reaching the space parameter with the maximum posterior probability ($a_{\rm max} \geq 300 \ \mu$m.)

\begin{figure*}
    \centering
    \includegraphics[width=1.0\textwidth]{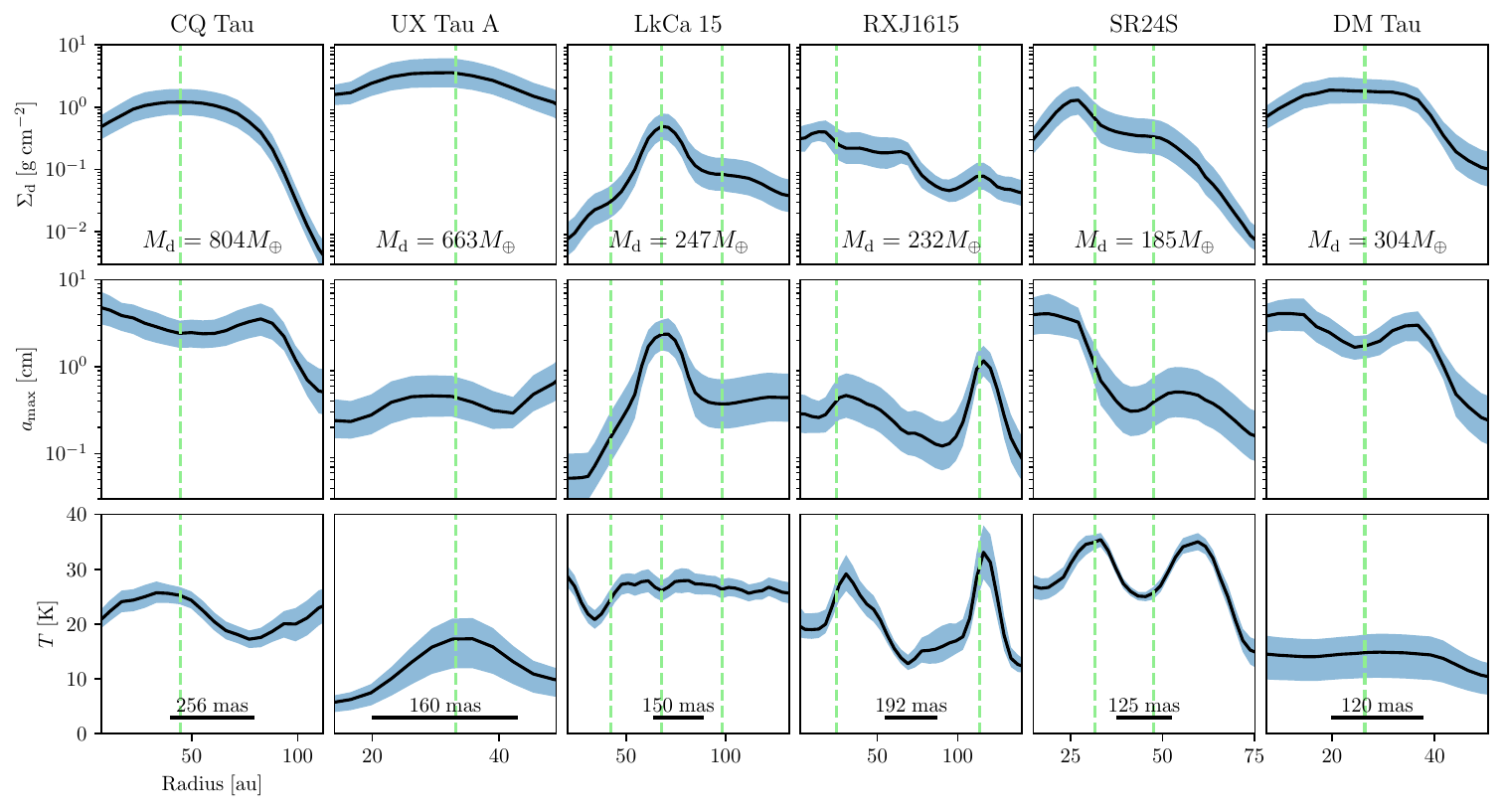} 
    \caption{Dust surface density (top panels), maximum grain size (middle panels), and dust temperature (bottom  panels) constraints from the SED fit. The vertical dashed lines mark the position of bright rings inferred at higher angular resolution. The inferred dust mass and angular resolution of each disk are shown in the bottom of the top and bottom panels, respectively.}
    \label{fig:Posterior}
\end{figure*}

The best fit of each parameter is shown in Figure \ref{fig:Posterior} for the 6 transition disks in this work. The vertical lines mark the position of the rings inferred from the visibility fit (Section \ref{sec:vismod}). The error bars are computed from the percentile 32 and 68 respectively. The dust mass is shown in the bottom of each top panel, while the angular resolution of each disk is shown in the bottom panels.

\begin{figure*}
    \centering
    \includegraphics[width=1.0\textwidth]{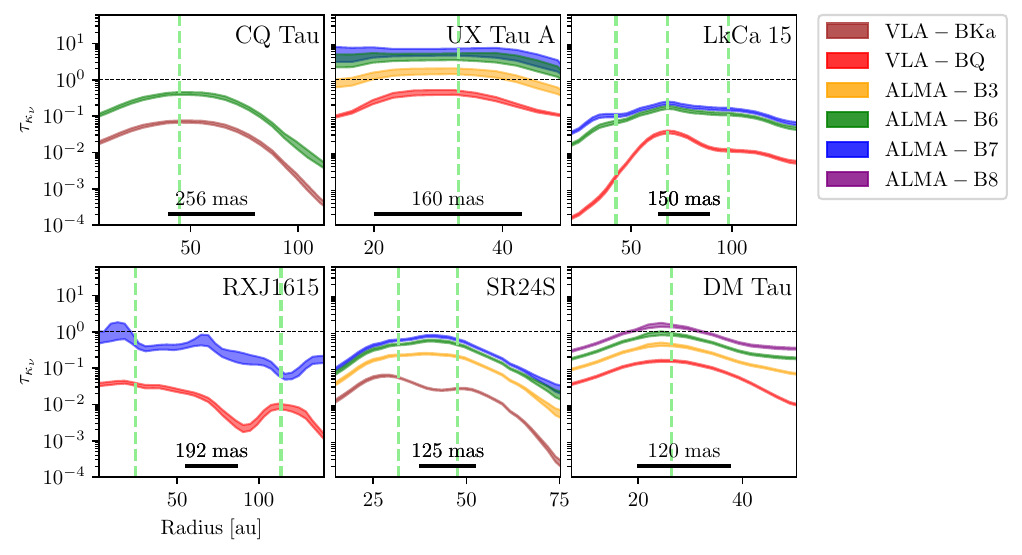} 
    \caption{Optical depth radial profiles computed from the SED fit at different wavelengths (see color code). The vertical dashed lines mark the position of bright rings inferred at higher angular resolution. The horizontal bar of each panel shows the angular resolution for each disk.}
    \label{fig:Optdepth}
\end{figure*}

Figure \ref{fig:Optdepth} shows the optical depth constraints resulting from the dust properties in Figure \ref{fig:Posterior} at different wavelengths.

\vspace{1cm}
\section{Discussion} \label{sec:discussion}
The results from the disk substructures computed from the visibility fit for each observation (Figures \ref{fig:Visibilities} and \ref{fig:Vis_RadProfile}), and the dust property constraints from the multi-wavelength fit (Section \ref{sec:multiwave}), give clues about the physical origin of the large cavities observed in the six transition disks in this work. However, before discussing the results from each individual disk (Sections \ref{sec:Planet}-\ref{sec:graingrowth}), we summarize the various caveats associated with our modeling below.

\subsection{Model limitations}
The exact dust properties inferred from the multi-wavelength analysis depend on the physical assumptions described in the previous section (e.g., particle size distribution slope, spherical and compact dust grains). However, general properties such as the peak position of the dust surface density or maximum grain size, are unchanged under these physical assumptions.

The maximum grain size estimations in this work are similar to previous analysis where the dust continuum SED has been modeled, and where millimeter and centimeter grain sizes have been inferred \citep[e.g.,][]{Perez_2012, Carrasco-Gonzalez_2019, Ueda_2020}. However, it is important to remark that smaller grain sizes (hundreds of microns or smaller grain sizes) have been inferred from polarized observations \citep[e.g.,][]{Kataoka_2016, Hull_2018, Mori_2019}. Our results are biased by the spectral index analysis. Alleviating the tension between the grain size estimations from polarization and spectral index is beyond the scope of this paper. An exception to the SED modeling has been found in the disk around DM Tau \cite{Liu_2024}, where the dust emission from localized substructures is consistent with $a_{\rm max} \gtrsim 300 \ \mu$m, but the rest of the disk is consistent with small grains ($\lesssim 50 \mu$m).

On the other hand, angular resolution limitations could impact the inferred dust properties profiles. In particular, the disk cavities could potentially be contaminated by emission from optically thick rings. In such cases, the cavities would present faint emission with low spectral indices, which can be interpreted as optically thin emission with large grain sizes. Given the angular resolution of the data, it is not possible to rule out the possibility of smearing effects within the cavities, specially in those disks where large grain sizes are inferred in this region.
A full discussion of the radial smearing effects are presented in Appendix \ref{app:smearing}. We also emphasize that our method for inferring the intrinsic angular resolution of a  \textsc{Frankenstein} fit does not assess the ability of such fit to resolve two independent structures (e.g., two rings). Rather, the intrinsic angular resolution corresponds to the ``point spread function'' of a \textsc{Frankenstein} fit done to an infinitely narrow ring. We tested different models with smaller separations between the two rings, but there is no clear limit at which the visibility fit is unable to separate them. The only limit is given by the sampling of the radial profile, which is not a physical measurement of the angular resolution, but is determined by the Nyquist sampling theorem \citep{Thompson_2017}, which should always be satisfied. See more details in Appendix \ref{app:angular}.

Additionally, we assume that the disks are axi-symmetric and compute their azimuthally averaged properties. However, some degree of asymmetry has been observed in some of them (see references in the discussion of each particular disk). In a few cases, these asymmetries are consistent with thermal noise (specially at long wavelengths).

Although we only study the multi-wavelength emission in a radial extend where the azimuthal asymmetries are low or negligible (See Section \ref{app:best_models}), the dust properties constraints in this work only pretend to explain the azimutally average dust properties of each analyzed disk, and a bi-dimensional analysis of the disks is beyond the scope of this work.
In some cases, such as DMTau, the axi-symmetric structure observed at ALMA Bands, and the azimuthal asymmetries observed at Band Q may be explained by optical depth effects \citep[see discussion in][]{Liu_2024}. In such cases, the constraints on the dust properties are mainly dominated by the morphology observed at ALMA wavelengths.

\subsection{ Disks with dust traps as a possible origin of the cavity}
\label{sec:Planet}

Disks with dust surface density and grain size maxima at ring positions are consistent with dust trap models \citep{Pinilla_2012b}. The dust trap can be induced by a companion (a planet or a close binary) or a dead zone.
However, if a deep gas cavity is also observed, it rules out the possibility of a dust trap created by a dead zone \citep{vanderMarel_2016}, and supports the idea of a companion origin.
In the following, we discuss the results obtained from disks where both a dust trap (identified in this work) and gas cavities (from the literature) have been inferred. These findings may suggest that the origin of these large cavities is a companion.

\vspace{0.3cm}
\textbf{UXTau A:} 
Dust continuum observations of the disk around UXTau A have been previously reported by \cite{Pinilla_2018} and \cite{Francis_2020}. From the visibility modeling in this work (Figure \ref{fig:Vis_RadProfile}), we confirm the ring-shaped structure, with a peak at $0.236^{\prime \prime}$ (33.0 au). Additionally, the disk emission exhibits azimuthal asymmetries at Band 6 and 7 (Figure \ref{fig:residuals}), with higher intensities observed towards the East. The Band 3 image appears non-axi-symmetric, but these asymmetries are created by the visibility coverage and the elongated beam, as shown by the perfect axi-symmetric disk model sampled at the same visibilities in Figure \ref{fig:models}. This is demonstrated in the model image of a perfectly axi-symmetric disk. Similarly, Band Q exhibits azimuthal structures along the major axis of the beam. The residuals of the enhanced emission at the South of the disk in Band Q has a SNR $< 5$, as shown in Figure \ref{fig:residuals}.

The dust properties constraints (Figure \ref{fig:Posterior}) shows a peak in the dust surface density, maximum grain size, and dust temperature aligning with the location of the main ring, exhibiting characteristics consistent with dust trap signatures, which can be triggered by a companion or a dead zone.

No planet candidates have been found in UXTau A using H$\alpha$ observations \citep{Zurlo_2020}. The sensitivity of the H$\alpha$ observations are consistent with a upper limit to the mass of the planet of $9$ Jupiter mass at $\sim$ 22 au. The $^{12}$CO (J:3-2) observations in \cite{Wolfer_2023} reveals a gas cavity and a ring structure in the gas that peaks near the dust ring. 

\vspace{0.3cm}
\textbf{LkCa15:}
Dust continuum observations of the disk around LkCa15 at high resolution have been studied by \cite{Isella_2014}, \cite{Facchini_2020} and \cite{Long_2022}, where three rings at at 0.277$^{\prime \prime}$ (43.5 au), 0.433$^{\prime \prime}$ (68.1 au), 0.636$^{\prime \prime}$ (100 au) have been inferred from the radial profiles in the image plane and from visibility modeling. We note a fourth faint outer ring at 1.070$^{\prime \prime}$ (168.2 au) from both the image plane and visibility modeling at ALMA wavelengths. This faint ring can also be inferred from the radial profile in the image plane at Band Q, but at a slightly smaller radius of 0.990$^{\prime \prime}$ (155.6 au).

The visibility modeling residuals at Band 6 and 7 in Figure \ref{fig:residuals} matches with those in \cite{Long_2022}. At Band Q, there are no residuals with SNR $> 5$. And at Band 9, there are positive/negative residuals in the far/near-side part of the disk, possibly due to an elevated emission surface at Band 9 \citep{Andrews_2021}, where the emission of small grains is traced.

The results from the SED fit in Figure \ref{fig:Posterior} for LkCa15 shows three local maxima for the dust surface density, aligning precisely with the radial positions of the three prominent rings observed in the dust continuum emission. The maximum grain size only has a clear peak around the main ring at $0.433^{\prime\prime}$ (68.1 au) at this resolution. These characteristics are indicative of dust trapping in the main ring, impeding the inward flow of dust grains and forming a large cavity.

No clear planetary signatures have been directly detected within the LkCa15 cavity, although planet candidates have been suggested \citep[e.g.,][]{Kraus_2012}. Comparing SPHERE/ALMA observations and the size of the observed gaps, \cite{Lodato_2019} and \cite{Asensio-Torres_2021} estimated that the upper limit of the mass of a planet companion within the cavity is $\lesssim 1$ Jupiter mass. On other side, the Lagrangian points identified by \cite{Long_2022} suggest the possible existence of a Neptune-mass planet positioned at a projected distance of 0.27$^{\prime \prime}$ (42.4 au). 

Finally, evidence of a gas cavity has been found in the molecular emission of $^{12}$CO (J=2-1), $^{13}$CO (J=2-1), and C$^{18}$O (J=2-1) \citep{Wolfer_2023}, suggesting that the origin of the cavity is a companion.

\vspace{0.3cm}
\textbf{RXJ1615:}
Millimeter dust continuum observations of the disk around RXJ1615 have been previously reported by \cite{vanderMarel_2015, Wolfer_2023}. The visibility modeling of the disk is shown in Figures \ref{fig:Visibilities}-\ref{fig:Vis_RadProfile}. At Band 9, the visibility modeling is sensitive to a ring at $0.82^{\prime \prime}$ (127.6 au), while a discernible dust cavity is absent, likely due to the limited angular resolution. The outer dust ring had previously been identified by \cite{Andrews_2011} and \cite{vanderMarel_2015}.

At Band 7, the radial profile shows a complex structure. Notably, it exhibits emission from various sources: a distinct compact inner disk, an asymmetric ring with a prominent peak at $0.158^{\prime \prime}$ (24.6 au), along with additional local peaks at approximately $0.251^{\prime \prime}$ (39.0 au) and $0.368^{\prime \prime}$ (57.3 au). Moreover, two further asymmetric peaks emerge at around $0.685^{\prime \prime}$ (106.6 au) and $0.839^{\prime \prime}$ (130.5 au), distinguishable in both the visibility fit and the image plane. Overall, the observations reveal the presence of up to 5 discernible rings and an inner disk.
In Band Q, the \textsc{Frankenstein} visibility fit fails to identify disk substructures, unlike the parametric fit, which successfully does. This is the only disk where the position of the outer ring is not fixed, but it is a free parameter (Table \ref{tab:parametric_fit}). The best-fit radial position for the outer ring is $0.73^{+0.05}_{-0.08}$, falling between the locations of the two outer rings observed in Band 7.

The dust surface density and maximum grain size (Figure \ref{fig:Posterior}) have clear local maxima at the outer ring, aligning with dust trap signatures. However, there is no clear evidence of a pressure trap in the inner ring (although it is suggested), due to the low angular resolution, which is not able to resolve the inner ring after convolution (beam size of $0.192^{\prime \prime}$ and a ring at $0.158^{\prime \prime}$, see Figure \ref{fig:Radial_Profiles}).

The findings in the outer ring align with dust trap signatures. On the other hand, a gas density drop has been detected within the disk cavity, as reported by \cite{vanderMarel_2015}, suggesting that the origin of the cavity is a companion. The mass of the companion candidate has been estimated as 4.5 Jupiter masses using SPHERE observations \citep{Asensio-Torres_2021}.

\subsection{Disks with possible evidence of enhanced grain growth and a pressure trap as the origin of the cavity} \label{sec:graingrowth}

In some of our transition disk sample, the inferred maximum grain size radial profile ($a_{\rm max}(r)$) monotonically decreases with radius, rather than showing a distinct drop within the cavity. Additionally, the distribution of $a_{\rm max}(r)$ appears relatively flat around the ring region, possibly indicating the presence of a dust trap.
On the other hand, the dust surface density radial profile shows a drop of mass within the cavity. The millimeter emission cavity in these cases is reproduced by a combination of very low densities and large dust grains with low opacity. As mentioned in Section \ref{sec:intro}, these characteristics are expected for cavities created by grain growth and a pressure bump.  Grain growth alone is insufficient to reproduce the low brightness within the cavity, and a pressure bump (created by a companion or a dead zone) is crucial for efficiently reducing the mass within the cavity \citep{Birnstiel_2012b}.

The absence of a gas gap within some of these disks further would suggest that the origin of the large cavity is not the same as in the rest of the disks. 
However, as mentioned at the beginning of this section, radial smearing of optically thick emission onto the cavity regions could affect the inferred dust properties. In particular, radial smearing could mimic large dust grain sizes within the cavity, as discussed in Appendix \ref{app:smearing}.
In the following, we discuss the results obtained from disks where a possible origin of the dust cavity may be attributed to low dust surface density triggered by a pressure bump in the ring, coupled with enhanced grain growth within the cavity, and where deep gas cavities have not been observed.

\vspace{0.3cm}
\textbf{CQTau:} 
The disk morphology of the dust continuum emission around CQTau has been previously reported at high angular resolution by \cite{Pinilla_2018} and  \cite{Ubeira_2019}. From the visibility modeling in this work (Figure \ref{fig:Vis_RadProfile}), we confirm the ring shape structure at Band 7, 6, and Ka, with a peak at $0.3^{\prime \prime}$ (44.8 au). The ring structure is also inferred from the image plane for Bands 6 and Ka. However, for Band 7, the angular resolution in the image plane smears the ring emission within the cavity, making it discernible only through the visibility fit.

According to \cite{Ubeira_2019}, the disk emission displays azimuthal asymmetries in both the continuum and CO isotopologue lines. These non-axi-symmetric structures are shown in Figure \ref{fig:residuals}. At Band 6, the disk emission appears brighter in the East and South-West regions compared to the average azimuthal values. A noticeable azimuthal asymmetry is also observed in the South-West of the disk at Band Ka (Figure \ref{fig:residuals}). However, this particular asymmetry has an SNR $< 5$.

The dust properties radial profiles for CQ Tau shows a dust surface density with a peak at the ring position. The maximum grain size monotonically decreases with radius, suggesting no strong evidence of a dust trap (at this angular resolution). However, a gas cavity has been inferred from the CO isotopologue emission in \cite{Ubeira_2019}. Their work shows that the gas cavity is consistent with a planet of 6-9 Jupiter masses located 20 au from the central star. Given the gas and dust observations, it is not clear what mechanism is able to explain the cavity observed in dust continuum emission for CQ Tau.

\vspace{0.3cm}
\textbf{SR24S:}
High angular resolution of the disk around SR24S at millimeter wavelengths have been previously reported by \cite{Pinilla_2017}, \cite{Cieza_2021} and \cite{ Francis_2020}.
The visibility modeling for SR24S (Figures \ref{fig:Visibilities}-\ref{fig:Vis_RadProfile}) shows an asymmetric ring structure at all wavelengths, similar to two adjacent ring structures, with peaks at $\sim 0.254^{\prime \prime}$ (29.0 au), $0.441^{\prime \prime}$ (50.3 au). The asymmetric ring has been already reported in previous works \citep[e.g.,][]{Cieza_2021}. The \textsc{Frankenstein} and parametric fit at Band Ka are able to resolve the two rings, at the same radial location observed in the ALMA data. 

The dust properties constraints for SR24S shows a dust surface density increasing towards the two rings, however the maximum grain size decreases with radius, and it only has a flat distribution around the ring at $0.254^{\prime \prime}$ (29.0 au). Large dust grain sizes are inferred within the cavity, suggesting that enhanced grain growth and a pressure bump are responsible of the deficit of millimeter emission in this region.

SR24S is the only disk among the transition disks studied by \cite{vanderMarel_2015} where a drop of at least 1 or 2 orders of magnitude in the gas surface density within the disk cavity is not inferred. \cite{Pinilla_2020} confirmed that the optically thin molecular emissions of $^{13}$CO and C$^{18}$O peak at the center of the disk cavity, indicating no gas deficit within the inner cavity of the disk, unlike many transition disks. As pointed out by \cite{Pinilla_2020}, a large cavity created by photoevaporation is unlikely due to the high accretion rate, the large dust cavity, and the gas emission within this region.

\vspace{0.3cm}
\textbf{DMTau:}
Millimeter dust continuum observations of the disk around DMTau at high resolution have been previously studied by many authors at several wavelengths, including \cite{Andrews_2011}, \cite{Semenov_2018}, \cite{Hashimoto_2021}, and \cite{Liu_2024}.

The observations presented by \cite{Hashimoto_2021} at high angular resolution reveal an inner ring ($<0.1^{\prime \prime}$, 14.4 au), a main ring with a peak at $\sim 0.16^{\prime \prime}$ (23.0 au), and two extended faint ring structures, as shown in the radial profile from the visibility fit at Band 6 in Figure \ref{fig:Vis_RadProfile}. The main ring at this wavelength only present weak azimuthal asymmetries, likely due optical depth effects. These asymmetries are better observed at optically thinner wavelengths using JVLA observations at Band Q \citep{Liu_2024}, where a visibility analysis shows that the Band Q intensity is azimuthally asymmetric on the spatial scale of $\sim$ 20 au or less. The spatial resolution in this work ($\sim 18$ au) is only marginally resolve those spatial scales.

The primary ring within the disk is consistently identified across all wavelengths in the visibility fit (ranging from Band 9 to Band Q). Faint outer structures are evident across all wavelengths, but their radial positions do not align with those observed at Band 6. These faint structures disappear when the radial profiles are convolved at the same angular resolution (Figure \ref{fig:Radial_Profiles}).

The results from the SED fit for DMTau in this work (Figure \ref{fig:Posterior}) shows a dust surface density increasing towards the position of the main ring. The maximum grain size decreases with radius, and it has a flat distribution close to the disk ring. Evidence of efficient grain growth in the inner most region of DMTau has been also found by \cite{Hashimoto_2021}. They found that the SED can be reproduced using a population of large dust grains (millimeter grains) that are less depleted than small dust grains (submicron size), consistent with efficient grain growth in the inner region of the disk, consistent with our grain size estimations within the cavity. However, the multi-wavelength analysis in \cite{Liu_2024} recently showed that the integrated SED from 8 to 700 GHz can be explained by two grain size components: one population of $a_{\rm max} \gtrsim 300$ microns sizes for the emission of small solid angles (as the knot observed in Band Q), and a second population of $a_{\rm max} <50$ microns for larger solid angles (similar to the spatial scale emission from ALMA wavelengths). These analysis and their apparent disagreement show why higher angular resolution observations where the emission from compact and extended sources can be resolved at several wavelengths.

No evidence of a gas gap has been found within the dust cavity \citep{Kudo_2018, Hashimoto_2021, vanderMarel_2021}, although the $^{12}$CO observations used in these works may not be able to resolve the gap, and the emission could be optically thick. \cite{Wolfer_2023} studied the spatial distribution of optically thinner molecular transitions in DMTau, including $^{13}$CO and C$^{18}$O, but no evidence of a deep gas cavity or spirals/arcs structure was found, although the angular resolution could also be insufficient. Therefore, the presence of a possible gas gap inside the dust cavity is still unknown.

As pointed out by \cite{Kudo_2018}, the origin of the dust cavity remains unclear. The high accretion rate disfavors photoevaporation, while the absence of a gas gap \citep{Wolfer_2023} disfavors a companion. In any case, a pressure bump at the ring location is needed to explain the deficit of dust mass within the cavity.

\section{Does the origin of a large cavity depend on the disk mass?}\label{sec:origin}

The constraints on dust properties shown in Figure \ref{fig:Posterior} offer insights into the origin of the observed large cavities. 
The results suggest that in UXTau A, LkCa15, and RXJ1615, the origin of the large cavities are dust traps created by a companion (a planet or a close binary). Conversely, in CQTau, SR24S and DMTau, a possible origin of the large cavities is a combination of monotonically increasing maximum grain size towards the disk center (which results in low emission at millimeter wavelengths), and a deficit of dust mass created by a pressure bump in the ring \citep{Birnstiel_2012b}. In the latter scenario, the pressure bump can be generated, for example, by a dead zone or a companion.

It is important to note that resolution effects may play a significant role in inferring the increasing maximum grain size towards the disk center, possibly due to the presence of inner disks, as shown in the brightness radial profile of SR24S at Band 3 or DMTau at Band 6 (Figure \ref{fig:Vis_RadProfile}). Although the inner disks vanish when convolved to a lower angular resolution (Figure \ref{fig:Radial_Profiles}), their emission could spread to outer radii within the cavity (see discussion in Appendix \ref{app:smearing}).


Table \ref{Tab:Origin} summarizes various physical mechanisms consistent with the observational signatures presented in this work. Additionally, it includes constraints on gas cavities, and the detection of spirals/arms in CO maps, which have been interpreted as potential evidence of embedded companions \citep{Wolfer_2023}. It is also important to remark that no evidence of a inner brown dwarf companions with a mass $>20 $M$_{\mathrm{Jup}}$ has been found in UXTau A, LkCa15, RXJ1615, and SR24S, while it is only tentative for DMTau \citep{Kraus_2011}.

The reported dust masses in Table \ref{Tab:Origin} are estimated by integrating the dust surface density profiles in Figure \ref{fig:Posterior}.
The final columns present similar observational signatures observed in two other transition disks (GMAur, Sz91), where multi-wavelength modeling has also been used to infer dust properties radial profiles.

\begin{table*}
\centering
\label{Tab:Origin}
\caption{Compatibility of the different large cavity origins and the observational signatures found in each disk.}
\begin{tabular}{c|cccccc|cc}
    \hline \hline
     & \multicolumn{6}{c|}{This work} & \multicolumn{2}{c}{Literature}\\
     &  CQTau & UXTau A & LkCa15 & RXJ1615 & SR24S & DMTau & GMAur & Sz91 \\
    \hline
    $M_* [M_{\odot}]$ & 1.63 & 1.4 & 1.32 & 1.1 & 0.87 & 0.39 & 1.1 & 0.54 \\
    $M_{\rm dust} [M_{\oplus}]$ &  804& 663 & 247  &  232 & 185 & 305 & 245 & 31.3 \\    
    Dust trap - companion & \pmark & \cmark & \cmark & \cmark & \pmark & \pmark & \pmark & \pmark \\    
    Dust trap - Dead zone & \pmark & \cmark & \cmark & \cmark & \pmark & \pmark & \pmark & \pmark\\    
    Grain growth &  \cmark$^{*}$ & \xmark & \xmark & \xmark & \cmark$^{*}$ & \cmark$^{*}$ & \cmark$^{*}$ & \cmark$^{*}$ \\
    Photo-evaporation & \xmark & \xmark & \xmark & \xmark & \xmark & \xmark & \xmark & \xmark \\    
    \hline
    Gas cavity$^{(1)}$ & \cmark & \cmark & \cmark & \cmark & \xmark & \xmark $^{**}$ & \cmark & \cmark \\    
    Kinematic spirals/arms$^{(2)}$ & \cmark & \cmark & \pmark & \cmark & - & \xmark & \xmark & \pmark \\
    \hline \hline
\end{tabular}
    \newline
    \cmark: Compatible, \xmark: No compatible, \pmark: Tentative. 
    $^{(*)}$Possibly affected by radial smearing of low angular resolution observations.
    $^{(1)}$Gas constraints from literature (references in Sections \ref{sec:Planet}-\ref{sec:graingrowth}), $^{(2)}$From kinematic signatures in \cite{Wolfer_2023}. $^{(**)}$Low angular resolution observations. Stellar mass references: GMAur \citep{Macias_2018}, Sz91 \citep{Francis_2020}, and those in Table \ref{tab:Stellar}
\end{table*}

The dust properties in the disks around GMAur \citep{Sierra_2021} and Sz91 \citep{Mauco_2021} show dust surface density maxima at the position of the rings, and a maximum grain size decreasing with radius, consistent with a large cavity created by a pressure bump and enhanced grain growth. However, radial smearing effects within the cavity cannot be ruled out given the spatial resolution of the observations. However, unlike SR24S and DMTau, gas cavities have been observed in GMAur and Sz91, and tentative kinematic spirals/arms in Sz91.

In our sample, the disk dust mass generally tends to increase with the stellar mass, as already reported by \cite{Pinilla_2020} in a large sample of disks. Note that the origin of the large cavities with the lowest stellar/dust mass in our sample is tentatively attributed a pressure bump and enhanced grain growth. Conversely, for higher stellar/dust masses, except possibly CQTau, the large cavities are consistent with dust traps formed by a companion. It is important to remark that the long wavelength data in this work (from Disks\@EVLA project, ID: AC982) are biased towards massive disks. Thus, a larger sample of transition disks covering lower disk masses is needed to confirm or rule out our suggestive findings, in addition to higher angular resolution, which can help rule out radial smearing effects.

It is known that disk mass can play a role in shaping the formation, evolution, and characteristics of protoplanetary disks \citep[e.g.,][and references therein]{Ribas_2015, Drkazkowska_2023}. In transition disks, the origin of the cavities could be mass dependent, as suggested by the results in this work. Further observational modeling, high angular resolution data, and simulations are necessary to delve deeper into exploring this possibility.

\section{Conclusions}\label{sec:conclusions}

We analyzed millimeter dust continuum observations of transition disks surrounding CQTau, UXTau A, LkCa15, RXJ1615, SR24S, and DMTau. A total of 26 observations were used, covering wavelengths from ALMA Band 9 (435 $\mu$m) to VLA Band Ka (8.8 mm).
These observations were used to fit the radial brightness profiles at sub-beam resolution, enabling a study of the radial distribution of solids. 

The constraints on the dust surface density, dust temperature, and maximum grain size radial profiles inferred from modeling the SED offer valuable insights regarding the physical origin of the observed large cavities in each disk. 
Our main conclusions about the disk morphology and origin of the large cavities in each disk are summarized as follows.

\begin{enumerate}
    \item The radial brightness profiles observed in the disks around CQTau and UXTau A reveal a Gaussian-like ring at all wavelengths, and no emission from an inner disk. Additionally, the point source at the longest VLA wavelengths is consistent with a notably low flux (upper limit of $\lesssim 7\ \mu$Jy) in both disks.
    
    \item Across multiple wavelengths, the brightness radial profiles observed in the disks around RXJ1615, LkCa15, SR24S, and DMTau display a multi-ring structure. Notably, RXJ1615 exhibits the highest number of discernible sub-structures, particularly evident at Band 7 where the high angular resolution enables the identification of up to 5 local maxima and an inner disk. In LkCa15 and SR24S, the number of rings or local maxima is 4 and 2, respectively.

    In the case of DMTau, the primary ring observed at the Band 6 wavelength displays asymmetry. Additionally, this profile shows two faint rings in the outer disk, which are poorly constrained across other wavelengths. The disks associated to RXJ1615, LkCa15, SR24S, and DMTau have an inner disk, which can be observed at the highest angular resolution observations. 

    \item The radial profiles of the dust surface density and maximum grain size in the disks around UXTau A, LkCa15, and RXJ1615 have peak values at the position of the main ring of each disk. These properties align with dust trap models, which may be formed by either a companion or a dead zone. Gas observations of these disks have unveiled gas cavities and kinematic spirals/arms, indicating that the formation of the dust trap and the large cavities in these disks could be attributed to a companion carving a deep gap.

    \item In CQTau, SR24S and DMTau, the dust surface density peaks at the ring positions, whereas the maximum grain size decreases with radius, displaying a plateau around the rings. Although the inferred large grain sizes within the cavity can be affected by radial smearing, these large grain sizes are able to reproduce the multi-wavelength observations.
    These findings suggest the possibility that large grains (with low opacity at millimeter wavelengths) are present in the cavity (unlike the other disks in this work), while the low dust surface densities constrained in the cavities might be triggered by a pressure bump in the ring (either by a companion or dead zone).
    No gas cavities and kinematic spirals/arms in molecular gas observations have been observed in SR24S and DMTau, suggesting that the origin of their cavities may differ from the rest of the disks in this work. However, the low angular resolution of existing dust continuum and molecular gas observations are not sufficient to favor a particular mechanism.

    \item The dust properties constraints in the transition disks in our sample may exhibit different characteristics, suggesting potentially different origins for the large cavities. In most of the massive disks (UXTau A, LkCa15, RXJ1615), the cavity formation can be attributed to a dust trap created by a companion. In contrast, for the less massive disks (SR24S, DMTau), the cavities may result from a deficit of dust mass created by a pressure bump and enhanced grain growth, leading to a reduction in the emergent intensity within the cavity. However, similar constraints are found in the massive disk around CQ Tau.
    Given the low angular resolution and possible radial smearing, the possibility of small grain sizes within the cavity (as those inferred from polarized observations) consistent with pressure traps, cannot be ruled out for CQTau, SR24S and DMTau.
\end{enumerate}

\acknowledgments
We are very thankful for the thoughtful suggestions of the anonymous referee that helped to improve our manuscript significantly.
A.S. acknowledges support from FONDECYT de Postdoctorado 2022 \#3220495. 
LP acknowledges support from ANID FONDECYT Regular \#1221442.
A.S, L.P., and M.B. acknowledge support from ANID BASAL project FB210003 and Programa de Cooperación Cientifica ECOS-ANID ECOS200049. 
We thank Nicolas Kurtovic for his help on the calculation of disk offsets and geometry constraints. We thank Elena Viscardi and Enrique Macias for their discussion about the effects of beam smearing and its impact on inferring dust properties.
This project has received funding from the European Research Council (ERC) under the European Union’s Horizon 2020 research and innovation programme (PROTOPLANETS, grant agreement No. 101002188).
T.H. acknowledges support from the European Research Council under the Horizon 2020 Framework Program via the ERC Advanced Grant Origins 83 24 28.

This paper makes use of the following ALMA data: 
ADS/JAO.ALMA \#2011.0.00320.S, 
ADS/JAO.ALMA \#2011.0.00742.S, 
ADS/JAO.ALMA \#2012.1.00870.S, 
ADS/JAO.ALMA \#2013.1.00091.S, 
ADS/JAO.ALMA \#2013.1.00157.S, 
ADS/JAO.ALMA \#2013.1.00498.S, 
ADS/JAO.ALMA \#2015.1.00296.S, 
ADS/JAO.ALMA \#2015.1.00888.S, 
ADS/JAO.ALMA \#2015.1.01137.S, 
ADS/JAO.ALMA \#2016.1.00565.S, 
ADS/JAO.ALMA \#2016.1.01042.S, 
ADS/JAO.ALMA \#2018.1.00350.S, 
ADS/JAO.ALMA \#2018.1.01255.S, 
ADS/JAO.ALMA \#2018.1.01755.S 

ALMA is a partnership of ESO (representing its member states), NSF (USA) and NINS (Japan), together with NRC (Canada), MOST and ASIAA (Taiwan), and KASI (Republic of Korea), in cooperation with the Republic of Chile. The Joint ALMA Observatory is operated by ESO, AUI/NRAO and NAOJ. In addition, publications from NA authors must include the standard NRAO acknowledgement: The National Radio Astronomy Observatory is a facility of the National Science Foundation operated under cooperative agreement by Associated Universities, Inc.

We acknowledge the use of data from the Very Large Array (VLA) AC982, 12B-196, 13B-381, operated by the National Radio Astronomy Observatory (NRAO). The National Radio Astronomy Observatory is a facility of the National Science Foundation operated under cooperative agreement by Associated Universities, Inc

In memory of Rosita Morales, who passed away during the course of this work.

\software{Astropy \citep{astropy:2013, astropy:2018}, CASA \citep{McMullin_2007}, Emcee \citep{Foreman_2013}, Frankenstein \citep{Jennings_2020}, Matplotlib \citep{Matplotlib_2007}, Numpy \citep{Numpy_2020}}.

\appendix

\section{Angular resolution from visibility modeling} \label{app:angular}
In Section \ref{sec:multiwave}, we fit the spectral energy distribution (SED) of the disks as a function of radius. This requires comparing the multi-wavelength radial profiles from the visibility modeling at the same angular resolution.
The uv-coverage of the observations in this work varies significantly. For instance, most of the Band 9 data were collected during ALMA Cycle 0 (in 2012), while some Band 3 or 6 datasets were observed after 2020, where the antenna configuration is more extended. Therefore, we anticipate differences in the intrinsic angular resolution of the visibility fits. As fully discussed in Section 3 in \cite{Jennings_2020}, the visibility fits with \textsc{Frankenstein} can attain sub-beam resolution.
To address the angular resolution problem, we test two different methodologies: Delta Ring, and Gaussian disk, which are described below.

\subsection{Delta Ring}
\label{subsec:delta_ring}
We generate synthetic visibility observations of an infinitely narrow ring (a delta function) centered at a radius of $r_0 = R_{\rm max}/2$ (where $R_{\rm max}$ denotes the maximum disk radius from the visibility fit\footnote{Note that $R_{\rm max}$ is not a rigorous measurement of the disk radius but rather the maximum radial extent in the visibility fit. As suggested by \cite{Jennings_2020}, a suitable choice for $R_{\rm max}$ is typically greater than 1.5 times the expected disk radius.}). The center $r_0$ was chosen to avoid boundary effects. However, negligible differences are found when using a different $r_0$. The synthetic visibilities of the delta ring $\delta (r-r_0)$ are computed by its Hankel transform $\mathcal{H}$ as
\begin{equation}
    \label{eq:Hankel_Delta}
    \mathcal{H} \left[ \frac{A_0}{2\pi r_0} \delta(r-r_0)\right] = A_0 J_0(2\pi r_0 q),
\end{equation}
where $A_0$ is a constant that represents the total flux, and $q$ are the de-projected baselines.

These visibilities (right hand of Equation \ref{eq:Hankel_Delta}) are sampled at the same uv-coverage of each data set, and then we use \textsc{Frankenstein} to fit them (using the same methodology and hyper-parameters described in Table \ref{tab:Visibility}) and try to retrieve the delta function's input brightness distribution.  \textsc{Frankenstein} reconstructs  a Gaussian function instead of a delta function, with a full width half maximum (FWHM) that depends on the uv-coverage. This FWHM represents the minimum spatial scale recoverable by the visibility fit, i.e., the intrinsic angular resolution of the visibility fit. A Gaussian function is fitted to the reconstructed brightness profile, and the FWHM is reported in the last column in Table \ref{tab:Visibility}.  We remark that this methodology is equivalent to the definition of the Point Spread Function (PSF), defined as the response of an imaging system (e.g. an interferometer) to a point source. Deconvolution and imaging software (e.g. \textsc{CASA}) fits the 2D intensity distribution of the PSF to quantify the minimum recoverable spatial scale for a given uv-coverage, in an analogous way to our constrain of the spatial scale that our method recovers when the source is an infinitely narrow ring.

\subsection{Gaussian Disk}
\label{subsec:gaussian_disk}
\begin{figure}
    \centering
    \includegraphics[width=\textwidth]{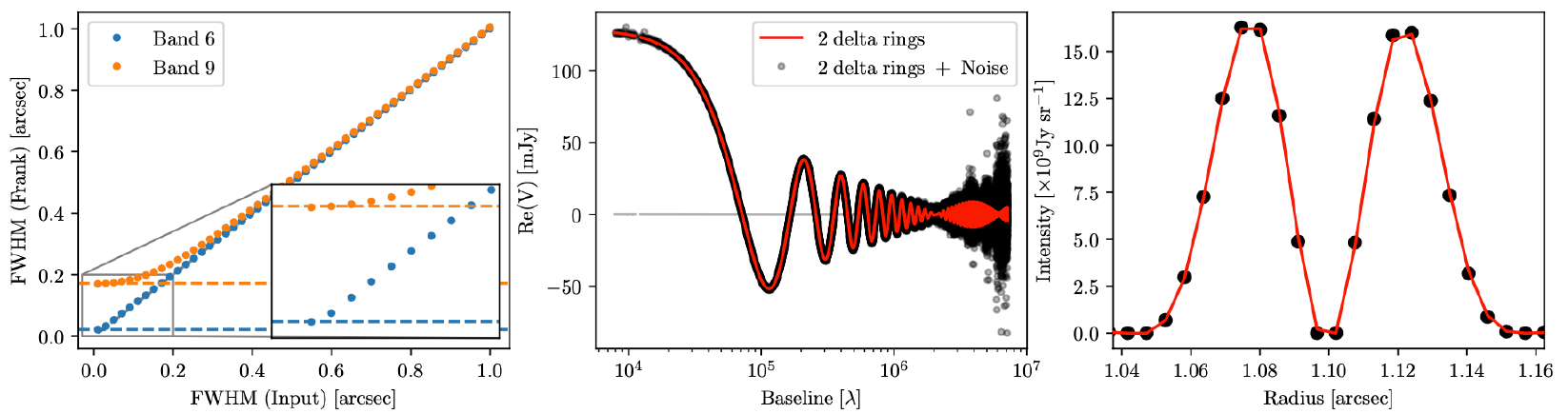}
    \caption{Left: Inferred FWHM of a gaussian disk as a function of the input FWHM model. The horizontal dashed line is the minimum recoverable spatial scale for a delta ring, which depends on the uv-coverage of each observation, such as the Band 6 (blue) and Band 9 (orange) observations of LkCa15. Middle: Real part of the visibilities as a function of baseline for two delta rings separated by 0.05$^{\prime \prime}$. The red and black dots represent a model without and with thermal noise, respectively. Right: Intensity radial profile obtained from \textsc{Frankenstein} for the two delta rings without (red) and with thermal noise (black).}
    \label{fig:Thermal_Noise}
\end{figure}

As a second method to infer the intrinsic angular resolution, we compute the visibilities of a full gaussian disk of a certain FWHM ($\Theta$) at the same uv-coverage of each observation. The visibilities are computed using the Hankel transform of a gaussian disk as
\begin{equation}
    \label{eq:Hankel_Gauss}
    \mathcal{H} \left[ \frac{1}{\Theta} \sqrt{\frac{4 \ln 2}{\pi}} \exp \left( -\frac{4 \ln2 r^2}{\Theta^2} \right) \right] = \exp \left(- \frac{(\pi \Theta q)^2}{4\ln2} \right).
\end{equation}
Then, these visibilities are fitted with \textsc{Frankenstein}, where a Gaussian function similar to that of the input model is obtained. The comparison between the input FWHM and that obtained with \textsc{Frankenstein} is shown in the left panel of Figure \ref{fig:Thermal_Noise} for the uv-coverage of LkCa15 at Band 6 and 9. Note that the FWHM obtained with \textsc{Frankenstein} differs from the input value when the latter tends to zero. The horizontal dashed lines are the angular resolutions computed from the delta rings in the previous section. These values represent the minimum spatial scale that a \textsc{Frankenstein} fit with a certain uv-coverage is able to resolve.

\subsection{The effects of thermal noise when resolving two delta rings}
The input models for a delta ring and Gaussian disk above were computed without including thermal noise to the visibilities. In this section we demonstrate that our visibility fit resolves the emission of two delta rings, with or without the inclusion of thermal noise.
The first input model consists of two delta rings separated by 50 mas around $R_{\rm max}/2 = 1.1^{\prime \prime}$. The visibilities of each delta ring are sampled at the uv-coverage of LkCa15 at Band 6 using the equation \ref{eq:Hankel_Delta}. The second model is the same but including thermal noise to the visibilities. The thermal noise is simulated by randomly choosing a visibility measurement within the visibilities error bar at each baseline. Then we use \textsc{Frankenstein} in both models to reconstruct the brightness radial profile. The visibility fit to the mock data are shown in the middle panel of Figure \ref{fig:Thermal_Noise}, and the reconstructed radial profiles are shown in the right panel of the same Figure. In both cases, the visibility fit is able to resolve both rings, with a FWHM of 21.5 mas, as already computed in Sections \ref{subsec:delta_ring}-\ref{subsec:gaussian_disk}.

\section{More on visibility modeling}
\subsection{Best Models and Residuals} \label{app:best_models}
\begin{figure*}
    \centering
    \includegraphics[width=\textwidth]{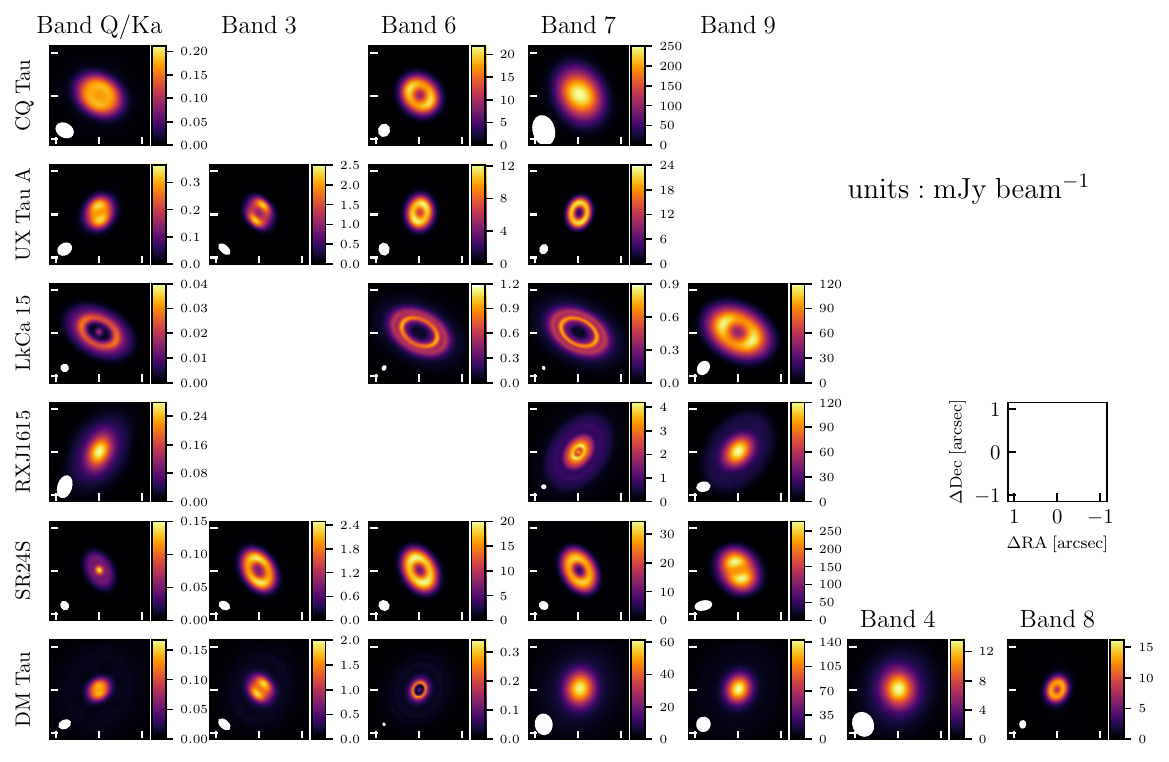}
    \caption{\textsc{clean} images of the disk models in our sample. The color bars are the same than in Figure \ref{fig:Observations}.}
    \label{fig:models}
\end{figure*}

The visibility models presented in Figure \ref{fig:Visibilities} are used to generate synthetic observations of the disks. Each model is sampled with the same uv-coverage as the observations and then imaged with identical observational parameters as described in Table \ref{tab:Observations}. Figure \ref{fig:models} shows the \textsc{clean} images of the disk models for all the observations. 

\begin{figure*}
    \centering
    \includegraphics[width=0.48\textwidth]{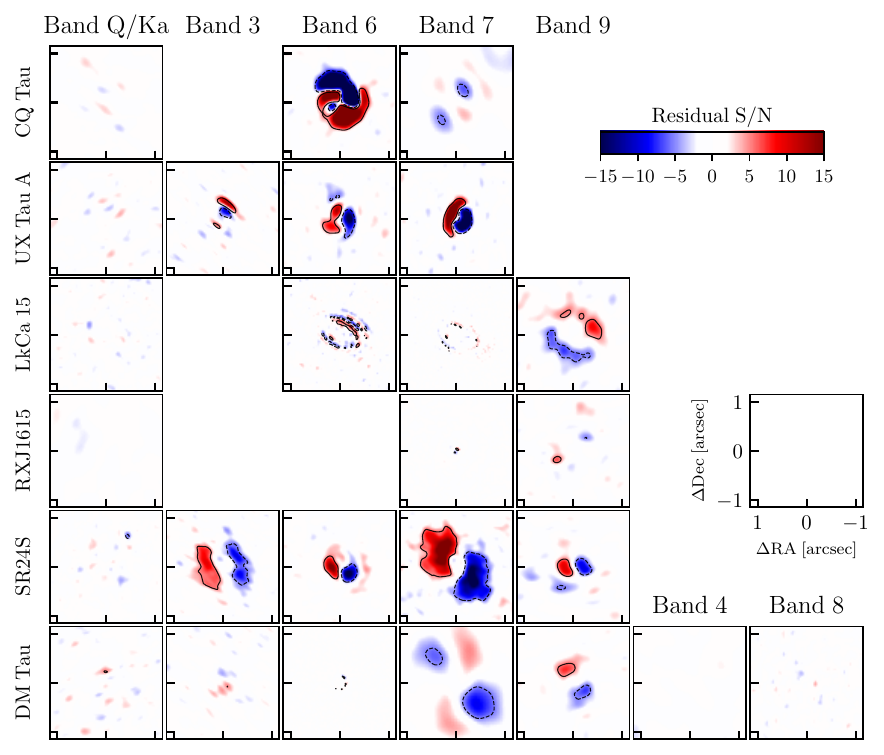}
    \includegraphics[width=0.48\textwidth]{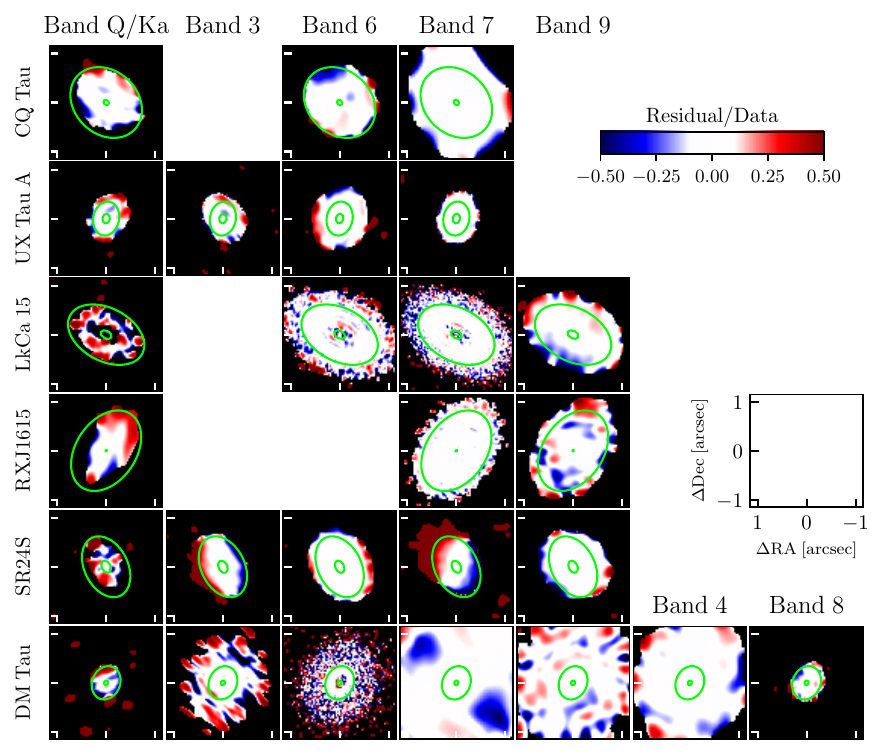}
    \caption{ Residual maps normalized with the rms (left) and the intensity maps (right).
    The name and band of the disk are shown in the top left and top right corner of each panel, respectively. The solid/dashed contours in the left figure are the region with a signal-to-noise (SNR) of +5/-5, respectively. The black area in the right figure is the region where the intensity map has a SNR $< 3$. The green ellipses show the minimum and maximum radius where dust properties constraints are computed in Figure \ref{fig:Posterior}.}
    \label{fig:residuals}
\end{figure*}

On the other hand, the visibility residuals (Data - Model) are computed and also used to generate synthetic observations. After sampling the visibility residuals with the same uv-coverage as the observations, they are imaged and shown in Figure \ref{fig:residuals}. Each map is normalized to the rms of each observation. The red/blue areas indicate regions with excess/deficit emission relative to the azimuthally averaged intensity of each observation.

\subsection{Parametric Models at VLA wavelengths} \label{app:parametric_models}
The parametric models for VLA data are given by a sum of Gaussian functions and a delta function $\delta(r)$ (mimic the compact emission from free-free) at the disk center:
\begin{equation}
    I(r) = \sum_i A_i \exp \left[ -\frac{(r-r_i)^2}{2 \sigma_i^2} \right] + f_0 \delta(r).
\end{equation}
The number of Gaussian functions for each disk is chosen based on the radial profile at ALMA wavelengths. The radial positions ($r_i$) of the Gaussian functions are fixed in all cases, determined by the ring locations observed at longer ALMA wavelengths for each respective disk.  We also conducted tests allowing $r_i$ to be a free parameter, and the resulting posterior probability distributions aligned with those when the radial positions were fixed. Therefore, to mitigate the ambiguity between the remaining free parameters, we opted to keep the radial positions of the Gaussian functions fixed.

The free parameters ($A_i, \sigma_i, f_0$) are determined via a Markov Chain Monte Carlo (MCMC) method \citep[implemented in the Python library \textsc{emcee},][]{Foreman_2013}, aiming to minimize the logarithm of the likelihood function given by Equation \ref{eq:posterior_vis}, and using 64 walkers and 10000 steps.
Table \ref{tab:parametric_fit} shows the parametric model chosen for each disk and the best fit for each parameter. Low and upper error bars correspond to percentiles 16 and 84, respectively.

The free-free flux (last column) for LkCa15, RXJ1615, and SR24S is 17.8, 44.7, and 11.0 $\mu$Jy respectively. These values are high compared with the rest of the disks in this work. \cite{Zapata_2017} showed that the SED of these disks can be described as a two-component power law: one of then associated to thermal dust emission, and the other to free-free emission, consistent with our estimations. Meanwhile, they claim that for UXTau A, a single power law (from optically thin dust emission) is enough to reproduce the SED. Our free-free flux estimation for UXTau A is only 0.05 $\mu$Jy, which is below the sensitivity (14.9 $\mu$Jy), and thus, it is not detected in this work.
For DM Tau, the best fit is consistent with a low free-free flux of only $0.02 \mu$ Jy, which is a non-detection given the sensitivity of $9 \mu$Jy. No free-free emission was detected in \cite{Zapata_2017} in DM Tau either. However, it is known that the free-free from ionized gas is variable \cite{Terada_2023}.
Finally, the free-free emission of CQ Tau is also low and below the sensitivity (7 $\mu$Jy). The non-detection of free-free emission is consistent with the lack of detection at 3.6 cm and 6.0 cm, suggesting that the flux at 7 mm is dominated by dust thermal emission \citep{Testi_2001}.

\begin{table}
    \centering 
    \caption{Best fits of the parametric model for VLA observations.}    
    \begin{tabular}{c|c|ccccccc}
        \hline \hline
       Source  & Parametric model & $\log_{10}(A_0)$  & fixed $r_0$ & $\log_{10} (\sigma_0)$ & $\log_{10}(A_1)$  & fixed $r_1$ & $\log_{10} (\sigma_1)$ & $\log_{10}(f_0)$ \\
               & & (Jy/sr) &  (arcsec) &  (arcsec) & (Jy/sr) & (arcsec) & (arcsec) & (Jy) \\
       \hline
       CQTau  & 1 Gaussian + point source & $7.83^{+0.02}_{-0.02}$  & 0.300 & $-0.92^{+0.02}_{-0.02}$ & - & - & - & $-6.29^{+1.05}_{-1.14}$ \\
       UXTau  A& 1 Gaussian + point source & $8.43^{+0.06}_{-0.05}$ & $0.236$ & $-1.26^{+0.06}_{-0.07}$ & - & - & - & $-7.31^{+1.85}_{-1.83}$ \\
       LkCa15 & 2 Gaussians + point source & $7.56^{+0.06}_{-0.06}$ & $0.433$ & $-1.21^{+0.07}_{-0.07}$ & $7.08^{+0.05}_{-0.06}$ & $0.635$ & $-0.81^{+0.06}_{-0.06}$ & $-4.75^{+0.10}_{-0.12}$\\
       RXJ1615 & 2 Gaussians + point source & $7.78^{+0.08}_{-0.08}$ & 0.158 & $-0.85^{+0.07}_{-0.09}$ & $7.52^{+1.09}_{-0.62}$ & $0.73^{+0.05}_{-0.08}$ & $-1.72^{+0.90}_{-1.18}$ & $-4.35^{+0.17}_{-0.29}$\\
       SR24S   & 2 Gaussian + point source & $7.74^{+0.11}_{-0.06}$ & $0.254$ & $-1.00^{+0.13}_{-0.21}$ & $7.42^{+1.25}_{-1.43}$ & $0.441$ & $-1.83^{+0.73}_{-0.77}$ & $-4.96^{+0.85}_{-0.97}$ \\       
       DMTau  & 2 Gaussian + point source & $8.26^{+0.12}_{-0.08}$ & $0.180$ & $-1.36^{+0.08}_{-0.13}$ & $6.57^{+0.53}_{-0.25}$ & $0.860$ & $-0.83^{+
       0.38}_{-0.66}$ & $-7.57^{+1.92}_{-1.68}$\\       
       \hline \hline
    \end{tabular}
    \label{tab:parametric_fit}
\end{table}

\section{Radial smearing effects}
\label{app:smearing}
The inferred dust properties depend on the lowest angular resolution of the multi-wavelength data. Higher contrasts between rings and gaps are expected in future high angular resolution observations. Radial smearing effects could potentially influence the inferred dust property profiles. For instance, optically thick emission at the rings could smear into the cavity region. This could mimic faint dust continuum emission with low spectral indices within the cavity, which might be attributed to large grain sizes (similar to those in CQTau, SR24 and DMTau). 

It is important to note that our methodology for studying the dust properties radial profiles cannot directly infer the deconvolved or convolved dust properties radial profiles. Instead, it provides only an approximation of those values, depending on the angular resolution. For example, in the optically thin regime and Rayleigh Jeans approximation, the emergent intensity is the product of three free parameters: $I_{\nu} \propto T_{\rm d} \Sigma_{\rm d} \kappa_{\nu}$. Since we observe the convolved intensities $\mathcal{C} (I_{\nu})$, we can only infer the convolution of the product of the three free parameters, rather than the product of individual convolved profiles $\mathcal{C} (T_{\rm d} \Sigma_{\rm d} \kappa_{\nu})  \neq \mathcal{C} (T_{\rm d}) \times \mathcal{C} (\Sigma_{\rm d}) \times \mathcal{C} (\kappa_{\nu})$. We only have good approximations to the convolved dust properties radial profiles when the angular resolution is sufficiently high (Viscardi et al. in prep).

\begin{figure}
    \centering
    \includegraphics{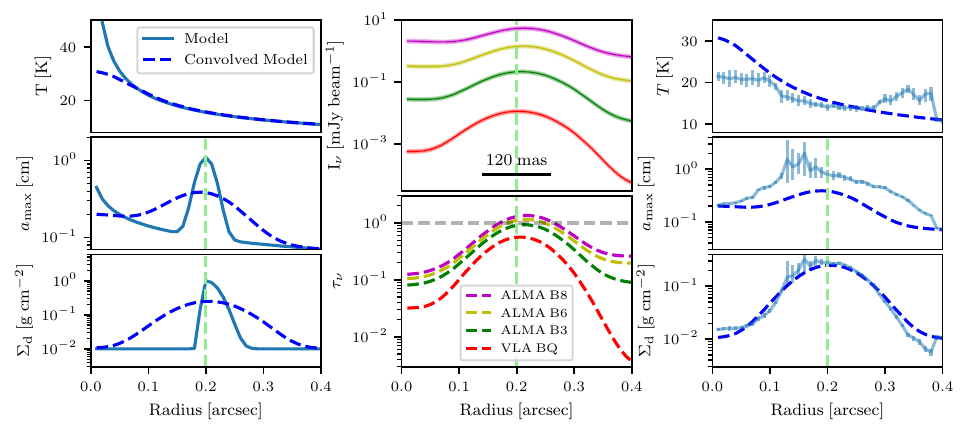}
    \caption{Left panels: Dust temperature, maximum grain size and dust surface density (from top to bottom) radial profile models (solid lines), and convolved profiles (dashed lines) with a 120 mas beam size (DMTau angular resolution). Middle panels: Intensity radial profiles (top) and optical depths (bottom) at the same resolution and multi-frequency observations as DMTau. Right panels: Dust temperature, maximum grain size, and dust surface density (from top to bottom) best fits (error bars), and input models (dashed lines).}
    \label{fig:smearing}
\end{figure}

We study the effects of radial smearing by creating a disk model with properties similar to those of DMTau, where large grain sizes are inferred within the disk cavity.
The temperature radial profile model is computed from a passively irradiated disk and the DMTau stellar parameters (Table \ref{tab:Stellar}). The maximum grain size model is given by $0.1(r/0.2^{\prime \prime})^{-0.5} + \exp[ - 0.5 ((r-0.2^{\prime \prime})/0.015^{\prime \prime})^2] $ cm, and the dust surface density is given by $0.01 + \exp(-0.5  (r-0.2^ {\prime \prime})^2/\sigma^2)$ g cm$^{-2}$, with $\sigma = 0.02^{\prime \prime}$, for $r> 0.2^{\prime \prime}$ and $\sigma = 0.005^{\prime \prime}$ for $r\leq 0.2^{\prime \prime}$ to mimic an asymmetric ring.
These radial profiles, chosen to represent dust trap signatures, are shown as a solid curve in the left panels of Figure \ref{fig:smearing}. The dashed lines in the same panels are the convolved radial profiles, using the resolution of the DMTau observations (120 mas). The middle panels show the convolved intensity profiles (top) and optical depths (bottom) corresponding to the same multi-wavelength observation as DMTau. The intensities and optical depths were originally computed using the dust properties models, and then these profiles were convolved to the same angular resolution.

The right panels show the best fit for dust temperature (top), maximum grain size (middle), and dust surface density (bottom). The blue dashed lines are the convolved radial profiles shown in the left panels. Any discrepancy between the radial profile from the posterior probability and the convolved radial profiles is due to the inability to independently infer the convolved dust properties radial profiles from observations (radial smearing effects), as discussed above. Note that the temperature and dust surface density radial profiles are well reproduced. However, the maximum grain size around the peak is radially smeared (much beyond the convolved radial profile), mimicking large grain sizes within the cavity, similar to the dust properties inferred in SR24S and DMTau (Figure \ref{fig:Posterior}). Thus, although large grain sizes within the cavity can reproduce the intensity profiles for CQTau, SR24S and DMTau, we cannot rule out the possibility of small grain sizes in this region, which are hidden by radial smearing effects.

\bibliography{main}{}

\begin{thebibliography}{}
\expandafter\ifx\csname natexlab\endcsname\relax\def\natexlab#1{#1}\fi
\providecommand{\url}[1]{\href{#1}{#1}}
\providecommand{\dodoi}[1]{doi:~\href{http://doi.org/#1}{\nolinkurl{#1}}}
\providecommand{\doeprint}[1]{\href{http://ascl.net/#1}{\nolinkurl{http://ascl.net/#1}}}
\providecommand{\doarXiv}[1]{\href{https://arxiv.org/abs/#1}{\nolinkurl{https://arxiv.org/abs/#1}}}

\bibitem[{{Alexander} {et~al.}(2006){Alexander}, {Clarke}, \&
  {Pringle}}]{Alexander_2006}
{Alexander}, R.~D., {Clarke}, C.~J., \& {Pringle}, J.~E. 2006, \mnras, 369,
  229, \dodoi{10.1111/j.1365-2966.2006.10294.x}

\bibitem[{{Andrews} {et~al.}(2011){Andrews}, {Wilner}, {Espaillat}, {Hughes},
  {Dullemond}, {McClure}, {Qi}, \& {Brown}}]{Andrews_2011}
{Andrews}, S.~M., {Wilner}, D.~J., {Espaillat}, C., {et~al.} 2011, \apj, 732,
  42, \dodoi{10.1088/0004-637X/732/1/42}

\bibitem[{{Andrews} {et~al.}(2009){Andrews}, {Wilner}, {Hughes}, {Qi}, \&
  {Dullemond}}]{Andrews_2009}
{Andrews}, S.~M., {Wilner}, D.~J., {Hughes}, A.~M., {Qi}, C., \& {Dullemond},
  C.~P. 2009, \apj, 700, 1502, \dodoi{10.1088/0004-637X/700/2/1502}

\bibitem[{{Andrews} {et~al.}(2014){Andrews}, {Chandler}, {Isella}, {Birnstiel},
  {Rosenfeld}, {Wilner}, {P{\'e}rez}, {Ricci}, {Carpenter}, {Calvet}, {Corder},
  {Deller}, {Dullemond}, {Greaves}, {Harris}, {Henning}, {Kwon}, {Lazio},
  {Linz}, {Mundy}, {Sargent}, {Storm}, \& {Testi}}]{Andrews_2014}
{Andrews}, S.~M., {Chandler}, C.~J., {Isella}, A., {et~al.} 2014, \apj, 787,
  148, \dodoi{10.1088/0004-637X/787/2/148}

\bibitem[{{Andrews} {et~al.}(2021){Andrews}, {Elder}, {Zhang}, {Huang},
  {Benisty}, {Kurtovic}, {Wilner}, {Zhu}, {Carpenter}, {P{\'e}rez}, {Teague},
  {Isella}, \& {Ricci}}]{Andrews_2021}
{Andrews}, S.~M., {Elder}, W., {Zhang}, S., {et~al.} 2021, \apj, 916, 51,
  \dodoi{10.3847/1538-4357/ac00b9}

\bibitem[{{Artymowicz} \& {Lubow}(1996)}]{Artymowicz_1996}
{Artymowicz}, P., \& {Lubow}, S.~H. 1996, \apjl, 467, L77,
  \dodoi{10.1086/310200}

\bibitem[{{Asensio-Torres} {et~al.}(2021){Asensio-Torres}, {Henning},
  {Cantalloube}, {Pinilla}, {Mesa}, {Garufi}, {Jorquera}, {Gratton}, {Chauvin},
  {Szul{\'a}gyi}, {van Boekel}, {Dong}, {Marleau}, {Benisty}, {Villenave},
  {Bergez-Casalou}, {Desgrange}, {Janson}, {Keppler}, {Langlois}, {M{\'e}nard},
  {Rickman}, {Stolker}, {Feldt}, {Fusco}, {Gluck}, {Pavlov}, \&
  {Ramos}}]{Asensio-Torres_2021}
{Asensio-Torres}, R., {Henning}, T., {Cantalloube}, F., {et~al.} 2021, \aap,
  652, A101, \dodoi{10.1051/0004-6361/202140325}

\bibitem[{{Astropy Collaboration} {et~al.}(2013){Astropy Collaboration},
  {Robitaille}, {Tollerud}, {Greenfield}, {Droettboom}, {Bray}, {Aldcroft},
  {Davis}, {Ginsburg}, {Price-Whelan}, {Kerzendorf}, {Conley}, {Crighton},
  {Barbary}, {Muna}, {Ferguson}, {Grollier}, {Parikh}, {Nair}, {Unther},
  {Deil}, {Woillez}, {Conseil}, {Kramer}, {Turner}, {Singer}, {Fox}, {Weaver},
  {Zabalza}, {Edwards}, {Azalee Bostroem}, {Burke}, {Casey}, {Crawford},
  {Dencheva}, {Ely}, {Jenness}, {Labrie}, {Lim}, {Pierfederici}, {Pontzen},
  {Ptak}, {Refsdal}, {Servillat}, \& {Streicher}}]{astropy:2013}
{Astropy Collaboration}, {Robitaille}, T.~P., {Tollerud}, E.~J., {et~al.} 2013,
  \aap, 558, A33, \dodoi{10.1051/0004-6361/201322068}

\bibitem[{{Astropy Collaboration} {et~al.}(2018){Astropy Collaboration},
  {Price-Whelan}, {Sip{\H{o}}cz}, {G{\"u}nther}, {Lim}, {Crawford}, {Conseil},
  {Shupe}, {Craig}, {Dencheva}, {Ginsburg}, {Vand erPlas}, {Bradley},
  {P{\'e}rez-Su{\'a}rez}, {de Val-Borro}, {Aldcroft}, {Cruz}, {Robitaille},
  {Tollerud}, {Ardelean}, {Babej}, {Bach}, {Bachetti}, {Bakanov}, {Bamford},
  {Barentsen}, {Barmby}, {Baumbach}, {Berry}, {Biscani}, {Boquien}, {Bostroem},
  {Bouma}, {Brammer}, {Bray}, {Breytenbach}, {Buddelmeijer}, {Burke},
  {Calderone}, {Cano Rodr{\'\i}guez}, {Cara}, {Cardoso}, {Cheedella}, {Copin},
  {Corrales}, {Crichton}, {D'Avella}, {Deil}, {Depagne}, {Dietrich}, {Donath},
  {Droettboom}, {Earl}, {Erben}, {Fabbro}, {Ferreira}, {Finethy}, {Fox},
  {Garrison}, {Gibbons}, {Goldstein}, {Gommers}, {Greco}, {Greenfield},
  {Groener}, {Grollier}, {Hagen}, {Hirst}, {Homeier}, {Horton}, {Hosseinzadeh},
  {Hu}, {Hunkeler}, {Ivezi{\'c}}, {Jain}, {Jenness}, {Kanarek}, {Kendrew},
  {Kern}, {Kerzendorf}, {Khvalko}, {King}, {Kirkby}, {Kulkarni}, {Kumar},
  {Lee}, {Lenz}, {Littlefair}, {Ma}, {Macleod}, {Mastropietro}, {McCully},
  {Montagnac}, {Morris}, {Mueller}, {Mumford}, {Muna}, {Murphy}, {Nelson},
  {Nguyen}, {Ninan}, {N{\"o}the}, {Ogaz}, {Oh}, {Parejko}, {Parley}, {Pascual},
  {Patil}, {Patil}, {Plunkett}, {Prochaska}, {Rastogi}, {Reddy Janga},
  {Sabater}, {Sakurikar}, {Seifert}, {Sherbert}, {Sherwood-Taylor}, {Shih},
  {Sick}, {Silbiger}, {Singanamalla}, {Singer}, {Sladen}, {Sooley},
  {Sornarajah}, {Streicher}, {Teuben}, {Thomas}, {Tremblay}, {Turner},
  {Terr{\'o}n}, {van Kerkwijk}, {de la Vega}, {Watkins}, {Weaver}, {Whitmore},
  {Woillez}, {Zabalza}, \& {Astropy Contributors}}]{astropy:2018}
{Astropy Collaboration}, {Price-Whelan}, A.~M., {Sip{\H{o}}cz}, B.~M., {et~al.}
  2018, \aj, 156, 123, \dodoi{10.3847/1538-3881/aabc4f}

\bibitem[{{Avenhaus} {et~al.}(2018){Avenhaus}, {Quanz}, {Garufi}, {Perez},
  {Casassus}, {Pinte}, {Bertrang}, {Caceres}, {Benisty}, \&
  {Dominik}}]{Avenhaus_2018}
{Avenhaus}, H., {Quanz}, S.~P., {Garufi}, A., {et~al.} 2018, \apj, 863, 44,
  \dodoi{10.3847/1538-4357/aab846}

\bibitem[{{Balbus} \& {Hawley}(1991)}]{Balbus_1991}
{Balbus}, S.~A., \& {Hawley}, J.~F. 1991, \apj, 376, 214,
  \dodoi{10.1086/170270}

\bibitem[{{Benisty} {et~al.}(2021){Benisty}, {Bae}, {Facchini}, {Keppler},
  {Teague}, {Isella}, {Kurtovic}, {P{\'e}rez}, {Sierra}, {Andrews},
  {Carpenter}, {Czekala}, {Dominik}, {Henning}, {Menard}, {Pinilla}, \&
  {Zurlo}}]{Benisty_2021}
{Benisty}, M., {Bae}, J., {Facchini}, S., {et~al.} 2021, \apjl, 916, L2,
  \dodoi{10.3847/2041-8213/ac0f83}

\bibitem[{{Birnstiel} {et~al.}(2012{\natexlab{a}}){Birnstiel}, {Andrews}, \&
  {Ercolano}}]{Birnstiel_2012b}
{Birnstiel}, T., {Andrews}, S.~M., \& {Ercolano}, B. 2012{\natexlab{a}}, \aap,
  544, A79, \dodoi{10.1051/0004-6361/201219262}

\bibitem[{{Birnstiel} {et~al.}(2016){Birnstiel}, {Fang}, \&
  {Johansen}}]{Birnstiel_2016}
{Birnstiel}, T., {Fang}, M., \& {Johansen}, A. 2016, \ssr, 205, 41,
  \dodoi{10.1007/s11214-016-0256-1}

\bibitem[{{Birnstiel} {et~al.}(2012{\natexlab{b}}){Birnstiel}, {Klahr}, \&
  {Ercolano}}]{Birnstiel_2012}
{Birnstiel}, T., {Klahr}, H., \& {Ercolano}, B. 2012{\natexlab{b}}, \aap, 539,
  A148, \dodoi{10.1051/0004-6361/201118136}

\bibitem[{{Birnstiel} {et~al.}(2018){Birnstiel}, {Dullemond}, {Zhu}, {Andrews},
  {Bai}, {Wilner}, {Carpenter}, {Huang}, {Isella}, {Benisty}, {P{\'e}rez}, \&
  {Zhang}}]{Birnstiel_2018}
{Birnstiel}, T., {Dullemond}, C.~P., {Zhu}, Z., {et~al.} 2018, \apjl, 869, L45,
  \dodoi{10.3847/2041-8213/aaf743}

\bibitem[{{Brown} {et~al.}(2009){Brown}, {Blake}, {Qi}, {Dullemond}, {Wilner},
  \& {Williams}}]{Brown_2009}
{Brown}, J.~M., {Blake}, G.~A., {Qi}, C., {et~al.} 2009, \apj, 704, 496,
  \dodoi{10.1088/0004-637X/704/1/496}

\bibitem[{{Brown} {et~al.}(2007){Brown}, {Blake}, {Dullemond}, {Mer{\'\i}n},
  {Augereau}, {Boogert}, {Evans}, {Geers}, {Lahuis}, {Kessler-Silacci},
  {Pontoppidan}, \& {van Dishoeck}}]{Brown_2007}
{Brown}, J.~M., {Blake}, G.~A., {Dullemond}, C.~P., {et~al.} 2007, \apjl, 664,
  L107, \dodoi{10.1086/520808}

\bibitem[{{Calvet} {et~al.}(2002){Calvet}, {D'Alessio}, {Hartmann}, {Wilner},
  {Walsh}, \& {Sitko}}]{Calvet_2002}
{Calvet}, N., {D'Alessio}, P., {Hartmann}, L., {et~al.} 2002, \apj, 568, 1008,
  \dodoi{10.1086/339061}

\bibitem[{{Carrasco-Gonz{\'a}lez} {et~al.}(2019){Carrasco-Gonz{\'a}lez},
  {Sierra}, {Flock}, {Zhu}, {Henning}, {Chandler}, {Galv{\'a}n-Madrid},
  {Mac{\'\i}as}, {Anglada}, {Linz}, {Osorio}, {Rodr{\'\i}guez}, {Testi},
  {Torrelles}, {P{\'e}rez}, \& {Liu}}]{Carrasco-Gonzalez_2019}
{Carrasco-Gonz{\'a}lez}, C., {Sierra}, A., {Flock}, M., {et~al.} 2019, \apj,
  883, 71, \dodoi{10.3847/1538-4357/ab3d33}

\bibitem[{{Casassus} {et~al.}(2013){Casassus}, {van der Plas}, {Perez}, {Dent},
  {Fomalont}, {Hagelberg}, {Hales}, {Jord{\'a}n}, {Mawet}, {M{\'e}nard},
  {Wootten}, {Wilner}, {Hughes}, {Schreiber}, {Girard}, {Ercolano}, {Canovas},
  {Rom{\'a}n}, \& {Salinas}}]{Casassus_2013}
{Casassus}, S., {van der Plas}, G.~M., {Perez}, S., {et~al.} 2013, \nat, 493,
  191, \dodoi{10.1038/nature11769}

\bibitem[{{Christiaens} {et~al.}(2024){Christiaens}, {Samland}, {Henning},
  {Portilla-Revelo}, {Perotti}, {Matthews}, {Absil}, {Decin}, {Kamp},
  {Boccaletti}, {Tabone}, {Marleau}, {van Dishoeck}, {G{\"u}del}, {Lagage},
  {Barrado}, {Garatti}, {Glauser}, {Olofsson}, {Ray}, {Scheithauer},
  {Vandenbussche}, {Waters}, {Arabhavi}, {Grant}, {Jang}, {Kanwar},
  {Schreiber}, {Schwarz}, {Temmink}, \& {{\"O}stlin}}]{Christiaens_2024}
{Christiaens}, V., {Samland}, M., {Henning}, T., {et~al.} 2024, arXiv e-prints,
  arXiv:2403.04855, \dodoi{10.48550/arXiv.2403.04855}

\bibitem[{{Cieza} {et~al.}(2021){Cieza}, {Gonz{\'a}lez-Ruilova}, {Hales},
  {Pinilla}, {Ru{\'\i}z-Rodr{\'\i}guez}, {Zurlo}, {Casassus}, {P{\'e}rez},
  {C{\'a}novas}, {Arce-Tord}, {Flock}, {Kurtovic}, {Marino}, {Nogueira},
  {Perez}, {Price}, {Principe}, \& {Williams}}]{Cieza_2021}
{Cieza}, L.~A., {Gonz{\'a}lez-Ruilova}, C., {Hales}, A.~S., {et~al.} 2021,
  \mnras, 501, 2934, \dodoi{10.1093/mnras/staa3787}

\bibitem[{{Delussu} {et~al.}(2024){Delussu}, {Birnstiel}, {Miotello},
  {Pinilla}, {Rosotti}, \& {Andrews}}]{Delussu_2024}
{Delussu}, L., {Birnstiel}, T., {Miotello}, A., {et~al.} 2024, arXiv e-prints,
  arXiv:2405.14501, \dodoi{10.48550/arXiv.2405.14501}

\bibitem[{{Dong} {et~al.}(2015){Dong}, {Zhu}, \& {Whitney}}]{Dong_2015}
{Dong}, R., {Zhu}, Z., \& {Whitney}, B. 2015, \apj, 809, 93,
  \dodoi{10.1088/0004-637X/809/1/93}

\bibitem[{{Draine}(2006)}]{Draine_2006}
{Draine}, B.~T. 2006, \apj, 636, 1114, \dodoi{10.1086/498130}

\bibitem[{{Dr{k{a}}{\.z}kowska} {et~al.}(2023){Dr{k{a}}{\.z}kowska}, {Bitsch},
  {Lambrechts}, {Mulders}, {Harsono}, {Vazan}, {Liu}, {Ormel}, {Kretke}, \&
  {Morbidelli}}]{Drkazkowska_2023}
{Dr{k{a}}{\.z}kowska}, J., {Bitsch}, B., {Lambrechts}, M., {et~al.} 2023, in
  Astronomical Society of the Pacific Conference Series, Vol. 534, Protostars
  and Planets VII, ed. S.~{Inutsuka}, Y.~{Aikawa}, T.~{Muto}, K.~{Tomida}, \&
  M.~{Tamura}, 717, \dodoi{10.48550/arXiv.2203.09759}

\bibitem[{{Dubrulle} {et~al.}(1995){Dubrulle}, {Morfill}, \&
  {Sterzik}}]{Dubrulle_1995}
{Dubrulle}, B., {Morfill}, G., \& {Sterzik}, M. 1995, \icarus, 114, 237,
  \dodoi{10.1006/icar.1995.1058}

\bibitem[{{Dullemond} {et~al.}(2001){Dullemond}, {Dominik}, \&
  {Natta}}]{Dullemond_2001}
{Dullemond}, C.~P., {Dominik}, C., \& {Natta}, A. 2001, \apj, 560, 957,
  \dodoi{10.1086/323057}

\bibitem[{{Espaillat} {et~al.}(2010){Espaillat}, {D'Alessio}, {Hern{\'a}ndez},
  {Nagel}, {Luhman}, {Watson}, {Calvet}, {Muzerolle}, \&
  {McClure}}]{Espaillat_2010}
{Espaillat}, C., {D'Alessio}, P., {Hern{\'a}ndez}, J., {et~al.} 2010, \apj,
  717, 441, \dodoi{10.1088/0004-637X/717/1/441}

\bibitem[{{Facchini} {et~al.}(2020){Facchini}, {Benisty}, {Bae}, {Loomis},
  {Perez}, {Ansdell}, {Mayama}, {Pinilla}, {Teague}, {Isella}, \&
  {Mann}}]{Facchini_2020}
{Facchini}, S., {Benisty}, M., {Bae}, J., {et~al.} 2020, \aap, 639, A121,
  \dodoi{10.1051/0004-6361/202038027}

\bibitem[{{Flock} {et~al.}(2015){Flock}, {Ruge}, {Dzyurkevich}, {Henning},
  {Klahr}, \& {Wolf}}]{Flock_2015}
{Flock}, M., {Ruge}, J.~P., {Dzyurkevich}, N., {et~al.} 2015, \aap, 574, A68,
  \dodoi{10.1051/0004-6361/201424693}

\bibitem[{{Foreman-Mackey} {et~al.}(2013){Foreman-Mackey}, {Hogg}, {Lang}, \&
  {Goodman}}]{Foreman_2013}
{Foreman-Mackey}, D., {Hogg}, D.~W., {Lang}, D., \& {Goodman}, J. 2013, \pasp,
  125, 306, \dodoi{10.1086/670067}

\bibitem[{{Francis} \& {van der Marel}(2020)}]{Francis_2020}
{Francis}, L., \& {van der Marel}, N. 2020, \apj, 892, 111,
  \dodoi{10.3847/1538-4357/ab7b63}

\bibitem[{{Gaia Collaboration}(2020)}]{Gaia_2020}
{Gaia Collaboration}. 2020, VizieR Online Data Catalog, I/350

\bibitem[{{G{\'a}rate} {et~al.}(2021){G{\'a}rate}, {Delage}, {Stadler},
  {Pinilla}, {Birnstiel}, {Stammler}, {Picogna}, {Ercolano}, {Franz}, \&
  {Lenz}}]{Garate_2021}
{G{\'a}rate}, M., {Delage}, T.~N., {Stadler}, J., {et~al.} 2021, \aap, 655,
  A18, \dodoi{10.1051/0004-6361/202141444}

\bibitem[{{Goldreich} \& {Tremaine}(1979)}]{Goldreich_1979}
{Goldreich}, P., \& {Tremaine}, S. 1979, \apj, 233, 857, \dodoi{10.1086/157448}

\bibitem[{Harris {et~al.}(2020)Harris, Millman, van~der Walt, Gommers,
  Virtanen, Cournapeau, Wieser, Taylor, Berg, Smith, Kern, Picus, Hoyer, van
  Kerkwijk, Brett, Haldane, del R{\'{i}}o, Wiebe, Peterson,
  G{\'{e}}rard-Marchant, Sheppard, Reddy, Weckesser, Abbasi, Gohlke, \&
  Oliphant}]{Numpy_2020}
Harris, C.~R., Millman, K.~J., van~der Walt, S.~J., {et~al.} 2020, Nature, 585,
  357, \dodoi{10.1038/s41586-020-2649-2}

\bibitem[{{Hartmann} {et~al.}(2016){Hartmann}, {Herczeg}, \&
  {Calvet}}]{Hartmann_2016}
{Hartmann}, L., {Herczeg}, G., \& {Calvet}, N. 2016, \araa, 54, 135,
  \dodoi{10.1146/annurev-astro-081915-023347}

\bibitem[{{Hashimoto} {et~al.}(2021){Hashimoto}, {Muto}, {Dong}, {Liu}, {van
  der Marel}, {Francis}, {Hasegawa}, \& {Tsukagoshi}}]{Hashimoto_2021}
{Hashimoto}, J., {Muto}, T., {Dong}, R., {et~al.} 2021, \apj, 911, 5,
  \dodoi{10.3847/1538-4357/abe59f}

\bibitem[{{Hayashi}(1981)}]{Hayashi_1981}
{Hayashi}, C. 1981, Progress of Theoretical Physics Supplement, 70, 35,
  \dodoi{10.1143/PTPS.70.35}

\bibitem[{{Hoang} {et~al.}(2018){Hoang}, {Lan}, {Vinh}, \& {Kim}}]{Hoang_2018}
{Hoang}, T., {Lan}, N.-Q., {Vinh}, N.-A., \& {Kim}, Y.-J. 2018, \apj, 862, 116,
  \dodoi{10.3847/1538-4357/aaccf0}

\bibitem[{{Huang} {et~al.}(2018){Huang}, {Andrews}, {Dullemond}, {Isella},
  {P{\'e}rez}, {Guzm{\'a}n}, {{\"O}berg}, {Zhu}, {Zhang}, {Bai}, {Benisty},
  {Birnstiel}, {Carpenter}, {Hughes}, {Ricci}, {Weaver}, \&
  {Wilner}}]{Huang_2018}
{Huang}, J., {Andrews}, S.~M., {Dullemond}, C.~P., {et~al.} 2018, \apjl, 869,
  L42, \dodoi{10.3847/2041-8213/aaf740}

\bibitem[{{Hull} {et~al.}(2018){Hull}, {Yang}, {Li}, {Kataoka}, {Stephens},
  {Andrews}, {Bai}, {Cleeves}, {Hughes}, {Looney}, {P{\'e}rez}, \&
  {Wilner}}]{Hull_2018}
{Hull}, C. L.~H., {Yang}, H., {Li}, Z.-Y., {et~al.} 2018, \apj, 860, 82,
  \dodoi{10.3847/1538-4357/aabfeb}

\bibitem[{Hunter(2007)}]{Matplotlib_2007}
Hunter, J.~D. 2007, Computing in Science \& Engineering, 9, 90,
  \dodoi{10.1109/MCSE.2007.55}

\bibitem[{{Isella} {et~al.}(2019{\natexlab{a}}){Isella}, {Benisty}, {Teague},
  {Bae}, {Keppler}, {Facchini}, \& {P{\'e}rez}}]{Isella_2019b}
{Isella}, A., {Benisty}, M., {Teague}, R., {et~al.} 2019{\natexlab{a}}, \apjl,
  879, L25, \dodoi{10.3847/2041-8213/ab2a12}

\bibitem[{{Isella} {et~al.}(2014){Isella}, {Chandler}, {Carpenter},
  {P{\'e}rez}, \& {Ricci}}]{Isella_2014}
{Isella}, A., {Chandler}, C.~J., {Carpenter}, J.~M., {P{\'e}rez}, L.~M., \&
  {Ricci}, L. 2014, \apj, 788, 129, \dodoi{10.1088/0004-637X/788/2/129}

\bibitem[{{Isella} {et~al.}(2019{\natexlab{b}}){Isella}, {Andrews},
  {Dullemond}, {P{\'e}rez}, {Huang}, {Birnstiel}, {Guzman}, {Kurtovic},
  {Zhang}, {Zhu}, {Bay}, {Benisty}, {Carpenter}, {Hughes}, {Oberg}, {Ricci},
  {Weaver}, \& {Wilner}}]{Isella_2019}
{Isella}, A., {Andrews}, S.~M., {Dullemond}, C.~P., {et~al.}
  2019{\natexlab{b}}, in 50th Annual Lunar and Planetary Science Conference,
  Lunar and Planetary Science Conference, 2821

\bibitem[{{Jennings} {et~al.}(2021){Jennings}, {Booth}, {Tazzari}, {Clarke}, \&
  {Rosotti}}]{Jennings_2021}
{Jennings}, J., {Booth}, R.~A., {Tazzari}, M., {Clarke}, C.~J., \& {Rosotti},
  G.~P. 2021, arXiv e-prints, arXiv:2103.02392.
\newblock \doarXiv{2103.02392}

\bibitem[{{Jennings} {et~al.}(2020){Jennings}, {Booth}, {Tazzari}, {Rosotti},
  \& {Clarke}}]{Jennings_2020}
{Jennings}, J., {Booth}, R.~A., {Tazzari}, M., {Rosotti}, G.~P., \& {Clarke},
  C.~J. 2020, \mnras, 495, 3209, \dodoi{10.1093/mnras/staa1365}

\bibitem[{{Kataoka} {et~al.}(2016{\natexlab{a}}){Kataoka}, {Muto}, {Momose},
  {Tsukagoshi}, \& {Dullemond}}]{Kataoka_2016b}
{Kataoka}, A., {Muto}, T., {Momose}, M., {Tsukagoshi}, T., \& {Dullemond},
  C.~P. 2016{\natexlab{a}}, \apj, 820, 54, \dodoi{10.3847/0004-637X/820/1/54}

\bibitem[{{Kataoka} {et~al.}(2016{\natexlab{b}}){Kataoka}, {Tsukagoshi},
  {Momose}, {Nagai}, {Muto}, {Dullemond}, {Pohl}, {Fukagawa}, {Shibai},
  {Hanawa}, \& {Murakawa}}]{Kataoka_2016}
{Kataoka}, A., {Tsukagoshi}, T., {Momose}, M., {et~al.} 2016{\natexlab{b}},
  \apjl, 831, L12, \dodoi{10.3847/2041-8205/831/2/L12}

\bibitem[{{Klahr} \& {Henning}(1997)}]{Klahr_1997}
{Klahr}, H.~H., \& {Henning}, T. 1997, \icarus, 128, 213,
  \dodoi{10.1006/icar.1997.5720}

\bibitem[{{Konigl} \& {Pudritz}(2000)}]{Konigl_2000}
{Konigl}, A., \& {Pudritz}, R.~E. 2000, in Protostars and Planets IV, ed.
  V.~{Mannings}, A.~P. {Boss}, \& S.~S. {Russell}, 759,
  \dodoi{10.48550/arXiv.astro-ph/9903168}

\bibitem[{{Kraus} \& {Ireland}(2012)}]{Kraus_2012}
{Kraus}, A.~L., \& {Ireland}, M.~J. 2012, \apj, 745, 5,
  \dodoi{10.1088/0004-637X/745/1/5}

\bibitem[{{Kraus} {et~al.}(2011){Kraus}, {Ireland}, {Martinache}, \&
  {Hillenbrand}}]{Kraus_2011}
{Kraus}, A.~L., {Ireland}, M.~J., {Martinache}, F., \& {Hillenbrand}, L.~A.
  2011, \apj, 731, 8, \dodoi{10.1088/0004-637X/731/1/8}

\bibitem[{{Kudo} {et~al.}(2018){Kudo}, {Hashimoto}, {Muto}, {Liu}, {Dong},
  {Hasegawa}, {Tsukagoshi}, \& {Konishi}}]{Kudo_2018}
{Kudo}, T., {Hashimoto}, J., {Muto}, T., {et~al.} 2018, \apjl, 868, L5,
  \dodoi{10.3847/2041-8213/aaeb1c}

\bibitem[{{Lin} \& {Papaloizou}(1979)}]{Lin_1979}
{Lin}, D.~N.~C., \& {Papaloizou}, J. 1979, \mnras, 186, 799,
  \dodoi{10.1093/mnras/186.4.799}

\bibitem[{{Lin} {et~al.}(2023){Lin}, {Li}, {Yang}, {Mu{\~n}oz}, {Looney},
  {Stephens}, {Hull}, {Fern{\'a}ndez-L{\'o}pez}, \& {Harrison}}]{Lin_2023}
{Lin}, Z.-Y.~D., {Li}, Z.-Y., {Yang}, H., {et~al.} 2023, \mnras, 520, 1210,
  \dodoi{10.1093/mnras/stad173}

\bibitem[{{Lin} {et~al.}(2024){Lin}, {Li}, {Stephens},
  {Fern{\'a}ndez-L{\'o}pez}, {Carrasco-Gonz{\'a}lez}, {Chandler}, {Pasetto},
  {Looney}, {Yang}, {Harrison}, {Sadavoy}, {Henning}, {Hughes}, {Kataoka},
  {Kwon}, {Muto}, \& {Segura-Cox}}]{Lin_2024}
{Lin}, Z.-Y.~D., {Li}, Z.-Y., {Stephens}, I.~W., {et~al.} 2024, \mnras, 528,
  843, \dodoi{10.1093/mnras/stae040}

\bibitem[{{Liu}(2019)}]{Liu_2019}
{Liu}, H.~B. 2019, \apjl, 877, L22, \dodoi{10.3847/2041-8213/ab1f8e}

\bibitem[{{Liu} {et~al.}(2024){Liu}, {Muto}, {Konishi}, {Chung}, {Hashimoto},
  {Doi}, {Dong}, {Kudo}, {Hasegawa}, {Terada}, \& {Kataoka}}]{Liu_2024}
{Liu}, H.~B., {Muto}, T., {Konishi}, M., {et~al.} 2024, \aap, 685, A18,
  \dodoi{10.1051/0004-6361/202348896}

\bibitem[{{Liu} {et~al.}(2017){Liu}, {Henning}, {Carrasco-Gonz{\'a}lez},
  {Chandler}, {Linz}, {Birnstiel}, {van Boekel}, {P{\'e}rez}, {Flock}, {Testi},
  {Rodr{\'\i}guez}, \& {Galv{\'a}n-Madrid}}]{Liu_2017}
{Liu}, Y., {Henning}, T., {Carrasco-Gonz{\'a}lez}, C., {et~al.} 2017, \aap,
  607, A74, \dodoi{10.1051/0004-6361/201629786}

\bibitem[{{Lodato} {et~al.}(2019){Lodato}, {Dipierro}, {Ragusa}, {Long},
  {Herczeg}, {Pascucci}, {Pinilla}, {Manara}, {Tazzari}, {Liu}, {Mulders},
  {Harsono}, {Boehler}, {M{\'e}nard}, {Johnstone}, {Salyk}, {van der Plas},
  {Cabrit}, {Edwards}, {Fischer}, {Hendler}, {Nisini}, {Rigliaco}, {Avenhaus},
  {Banzatti}, \& {Gully-Santiago}}]{Lodato_2019}
{Lodato}, G., {Dipierro}, G., {Ragusa}, E., {et~al.} 2019, \mnras, 486, 453,
  \dodoi{10.1093/mnras/stz913}

\bibitem[{{Long} {et~al.}(2022){Long}, {Andrews}, {Zhang}, {Qi}, {Benisty},
  {Facchini}, {Isella}, {Wilner}, {Bae}, {Huang}, {Loomis}, {{\"O}berg}, \&
  {Zhu}}]{Long_2022}
{Long}, F., {Andrews}, S.~M., {Zhang}, S., {et~al.} 2022, \apjl, 937, L1,
  \dodoi{10.3847/2041-8213/ac8b10}

\bibitem[{{Mac{\'\i}as} {et~al.}(2021){Mac{\'\i}as}, {Guerra-Alvarado},
  {Carrasco-Gonz{\'a}lez}, {Ribas}, {Espaillat}, {Huang}, \&
  {Andrews}}]{Macias_2021}
{Mac{\'\i}as}, E., {Guerra-Alvarado}, O., {Carrasco-Gonz{\'a}lez}, C., {et~al.}
  2021, \aap, 648, A33, \dodoi{10.1051/0004-6361/202039812}

\bibitem[{{Mac{\'\i}as} {et~al.}(2018){Mac{\'\i}as}, {Espaillat}, {Ribas},
  {Schwarz}, {Anglada}, {Osorio}, {Carrasco-Gonz{\'a}lez}, {G{\'o}mez}, \&
  {Robinson}}]{Macias_2018}
{Mac{\'\i}as}, E., {Espaillat}, C.~C., {Ribas}, {\'A}., {et~al.} 2018, \apj,
  865, 37, \dodoi{10.3847/1538-4357/aad811}

\bibitem[{{Manara} {et~al.}(2014){Manara}, {Testi}, {Natta}, {Rosotti},
  {Benisty}, {Ercolano}, \& {Ricci}}]{Manara_2014}
{Manara}, C.~F., {Testi}, L., {Natta}, A., {et~al.} 2014, \aap, 568, A18,
  \dodoi{10.1051/0004-6361/201323318}

\bibitem[{{Mauc{\'o}} {et~al.}(2021){Mauc{\'o}}, {Carrasco-Gonz{\'a}lez},
  {Schreiber}, {Sierra}, {Olofsson}, {Bayo}, {Caceres}, {Canovas}, \&
  {Palau}}]{Mauco_2021}
{Mauc{\'o}}, K., {Carrasco-Gonz{\'a}lez}, C., {Schreiber}, M.~R., {et~al.}
  2021, \apj, 923, 128, \dodoi{10.3847/1538-4357/ac21d0}

\bibitem[{{McMullin} {et~al.}(2007){McMullin}, {Waters}, {Schiebel}, {Young},
  \& {Golap}}]{McMullin_2007}
{McMullin}, J.~P., {Waters}, B., {Schiebel}, D., {Young}, W., \& {Golap}, K.
  2007, in Astronomical Society of the Pacific Conference Series, Vol. 376,
  Astronomical Data Analysis Software and Systems XVI, ed. R.~A. {Shaw},
  F.~{Hill}, \& D.~J. {Bell}, 127

\bibitem[{{Mori} {et~al.}(2019){Mori}, {Kataoka}, {Ohashi}, {Momose}, {Muto},
  {Nagai}, \& {Tsukagoshi}}]{Mori_2019}
{Mori}, T., {Kataoka}, A., {Ohashi}, S., {et~al.} 2019, \apj, 883, 16,
  \dodoi{10.3847/1538-4357/ab3575}

\bibitem[{{Mouillet} {et~al.}(2001){Mouillet}, {Lagrange}, {Augereau}, \&
  {M{\'e}nard}}]{Mouillet_2001}
{Mouillet}, D., {Lagrange}, A.~M., {Augereau}, J.~C., \& {M{\'e}nard}, F. 2001,
  \aap, 372, L61, \dodoi{10.1051/0004-6361:20010660}

\bibitem[{{Muto} {et~al.}(2012){Muto}, {Grady}, {Hashimoto}, {Fukagawa},
  {Hornbeck}, {Sitko}, {Russell}, {Werren}, {Cur{\'e}}, {Currie}, {Ohashi},
  {Okamoto}, {Momose}, {Honda}, {Inutsuka}, {Takeuchi}, {Dong}, {Abe},
  {Brandner}, {Brandt}, {Carson}, {Egner}, {Feldt}, {Fukue}, {Goto}, {Guyon},
  {Hayano}, {Hayashi}, {Hayashi}, {Henning}, {Hodapp}, {Ishii}, {Iye},
  {Janson}, {Kandori}, {Knapp}, {Kudo}, {Kusakabe}, {Kuzuhara}, {Matsuo},
  {Mayama}, {McElwain}, {Miyama}, {Morino}, {Moro-Martin}, {Nishimura}, {Pyo},
  {Serabyn}, {Suto}, {Suzuki}, {Takami}, {Takato}, {Terada}, {Thalmann},
  {Tomono}, {Turner}, {Watanabe}, {Wisniewski}, {Yamada}, {Takami}, {Usuda}, \&
  {Tamura}}]{Muto_2012}
{Muto}, T., {Grady}, C.~A., {Hashimoto}, J., {et~al.} 2012, \apjl, 748, L22,
  \dodoi{10.1088/2041-8205/748/2/L22}

\bibitem[{{Natta} {et~al.}(2006){Natta}, {Testi}, \& {Randich}}]{Natta_2006}
{Natta}, A., {Testi}, L., \& {Randich}, S. 2006, \aap, 452, 245,
  \dodoi{10.1051/0004-6361:20054706}

\bibitem[{{Paneque-Carre{\~n}o} {et~al.}(2021){Paneque-Carre{\~n}o},
  {P{\'e}rez}, {Benisty}, {Hall}, {Veronesi}, {Lodato}, {Sierra}, {Carpenter},
  {Andrews}, {Bae}, {Henning}, {Kwon}, {Linz}, {Loinard}, {Pinte}, {Ricci},
  {Tazzari}, {Testi}, \& {Wilner}}]{Paneque_2021}
{Paneque-Carre{\~n}o}, T., {P{\'e}rez}, L.~M., {Benisty}, M., {et~al.} 2021,
  \apj, 914, 88, \dodoi{10.3847/1538-4357/abf243}

\bibitem[{{Pascucci} {et~al.}(2023){Pascucci}, {Cabrit}, {Edwards}, {Gorti},
  {Gressel}, \& {Suzuki}}]{Pascucci_2023}
{Pascucci}, I., {Cabrit}, S., {Edwards}, S., {et~al.} 2023, in Astronomical
  Society of the Pacific Conference Series, Vol. 534, Protostars and Planets
  VII, ed. S.~{Inutsuka}, Y.~{Aikawa}, T.~{Muto}, K.~{Tomida}, \& M.~{Tamura},
  567, \dodoi{10.48550/arXiv.2203.10068}

\bibitem[{{Pearson}(1999)}]{Pearson_1999}
{Pearson}, T.~J. 1999, in Astronomical Society of the Pacific Conference
  Series, Vol. 180, Synthesis Imaging in Radio Astronomy II, ed. G.~B.
  {Taylor}, C.~L. {Carilli}, \& R.~A. {Perley}, 335

\bibitem[{{P{\'e}rez} {et~al.}(2014){P{\'e}rez}, {Isella}, {Carpenter}, \&
  {Chandler}}]{Perez_2014}
{P{\'e}rez}, L.~M., {Isella}, A., {Carpenter}, J.~M., \& {Chandler}, C.~J.
  2014, \apjl, 783, L13, \dodoi{10.1088/2041-8205/783/1/L13}

\bibitem[{{P{\'e}rez} {et~al.}(2012){P{\'e}rez}, {Carpenter}, {Chandler},
  {Isella}, {Andrews}, {Ricci}, {Calvet}, {Corder}, {Deller}, {Dullemond},
  {Greaves}, {Harris}, {Henning}, {Kwon}, {Lazio}, {Linz}, {Mundy}, {Sargent},
  {Storm}, {Testi}, \& {Wilner}}]{Perez_2012}
{P{\'e}rez}, L.~M., {Carpenter}, J.~M., {Chandler}, C.~J., {et~al.} 2012,
  \apjl, 760, L17, \dodoi{10.1088/2041-8205/760/1/L17}

\bibitem[{{P{\'e}rez} {et~al.}(2015){P{\'e}rez}, {Chandler}, {Isella},
  {Carpenter}, {Andrews}, {Calvet}, {Corder}, {Deller}, {Dullemond}, {Greaves},
  {Harris}, {Henning}, {Kwon}, {Lazio}, {Linz}, {Mundy}, {Ricci}, {Sargent},
  {Storm}, {Tazzari}, {Testi}, \& {Wilner}}]{Perez_2015}
{P{\'e}rez}, L.~M., {Chandler}, C.~J., {Isella}, A., {et~al.} 2015, \apj, 813,
  41, \dodoi{10.1088/0004-637X/813/1/41}

\bibitem[{{P{\'e}rez} {et~al.}(2016){P{\'e}rez}, {Carpenter}, {Andrews},
  {Ricci}, {Isella}, {Linz}, {Sargent}, {Wilner}, {Henning}, {Deller},
  {Chandler}, {Dullemond}, {Lazio}, {Menten}, {Corder}, {Storm}, {Testi},
  {Tazzari}, {Kwon}, {Calvet}, {Greaves}, {Harris}, \& {Mundy}}]{Perez_2016}
{P{\'e}rez}, L.~M., {Carpenter}, J.~M., {Andrews}, S.~M., {et~al.} 2016,
  Science, 353, 1519, \dodoi{10.1126/science.aaf8296}

\bibitem[{{Pinilla} {et~al.}(2012{\natexlab{a}}){Pinilla}, {Benisty}, \&
  {Birnstiel}}]{Pinilla_2012}
{Pinilla}, P., {Benisty}, M., \& {Birnstiel}, T. 2012{\natexlab{a}}, \aap, 545,
  A81, \dodoi{10.1051/0004-6361/201219315}

\bibitem[{{Pinilla} {et~al.}(2012{\natexlab{b}}){Pinilla}, {Birnstiel},
  {Ricci}, {Dullemond}, {Uribe}, {Testi}, \& {Natta}}]{Pinilla_2012b}
{Pinilla}, P., {Birnstiel}, T., {Ricci}, L., {et~al.} 2012{\natexlab{b}}, \aap,
  538, A114, \dodoi{10.1051/0004-6361/201118204}

\bibitem[{{Pinilla} {et~al.}(2016){Pinilla}, {Flock}, {Ovelar}, \&
  {Birnstiel}}]{Pinilla_2016}
{Pinilla}, P., {Flock}, M., {Ovelar}, M. d.~J., \& {Birnstiel}, T. 2016, \aap,
  596, A81, \dodoi{10.1051/0004-6361/201628441}

\bibitem[{{Pinilla} {et~al.}(2020){Pinilla}, {Pascucci}, \&
  {Marino}}]{Pinilla_2020}
{Pinilla}, P., {Pascucci}, I., \& {Marino}, S. 2020, \aap, 635, A105,
  \dodoi{10.1051/0004-6361/201937003}

\bibitem[{{Pinilla} {et~al.}(2014){Pinilla}, {Benisty}, {Birnstiel}, {Ricci},
  {Isella}, {Natta}, {Dullemond}, {Quiroga-Nu{\~n}ez}, {Henning}, \&
  {Testi}}]{Pinilla_2014}
{Pinilla}, P., {Benisty}, M., {Birnstiel}, T., {et~al.} 2014, \aap, 564, A51,
  \dodoi{10.1051/0004-6361/201323322}

\bibitem[{{Pinilla} {et~al.}(2017){Pinilla}, {P{\'e}rez}, {Andrews}, {van der
  Marel}, {van Dishoeck}, {Ataiee}, {Benisty}, {Birnstiel}, {Juh{\'a}sz},
  {Natta}, {Ricci}, \& {Testi}}]{Pinilla_2017}
{Pinilla}, P., {P{\'e}rez}, L.~M., {Andrews}, S., {et~al.} 2017, \apj, 839, 99,
  \dodoi{10.3847/1538-4357/aa6973}

\bibitem[{{Pinilla} {et~al.}(2018){Pinilla}, {Tazzari}, {Pascucci}, {Youdin},
  {Garufi}, {Manara}, {Testi}, {van der Plas}, {Barenfeld}, {Canovas}, {Cox},
  {Hendler}, {P{\'e}rez}, \& {van der Marel}}]{Pinilla_2018}
{Pinilla}, P., {Tazzari}, M., {Pascucci}, I., {et~al.} 2018, \apj, 859, 32,
  \dodoi{10.3847/1538-4357/aabf94}

\bibitem[{{Reg{\'a}ly} {et~al.}(2012){Reg{\'a}ly}, {Juh{\'a}sz}, {S{\'a}ndor},
  \& {Dullemond}}]{Regaly_2012}
{Reg{\'a}ly}, Z., {Juh{\'a}sz}, A., {S{\'a}ndor}, Z., \& {Dullemond}, C.~P.
  2012, \mnras, 419, 1701, \dodoi{10.1111/j.1365-2966.2011.19834.x}

\bibitem[{{Ribas} {et~al.}(2015){Ribas}, {Bouy}, \& {Mer{\'\i}n}}]{Ribas_2015}
{Ribas}, {\'A}., {Bouy}, H., \& {Mer{\'\i}n}, B. 2015, \aap, 576, A52,
  \dodoi{10.1051/0004-6361/201424846}

\bibitem[{{Rich} {et~al.}(2021){Rich}, {Teague}, {Monnier}, {Davies}, {Bosman},
  {Harries}, {Calvet}, {Adams}, {Wilner}, \& {Zhu}}]{Rich_2021}
{Rich}, E.~A., {Teague}, R., {Monnier}, J.~D., {et~al.} 2021, \apj, 913, 138,
  \dodoi{10.3847/1538-4357/abf92e}

\bibitem[{{Rilinger} {et~al.}(2023){Rilinger}, {Espaillat}, {Xin}, {Ribas},
  {Mac{\'\i}as}, \& {Luettgen}}]{Rilinger_2023}
{Rilinger}, A.~M., {Espaillat}, C.~C., {Xin}, Z., {et~al.} 2023, \apj, 944, 66,
  \dodoi{10.3847/1538-4357/aca905}

\bibitem[{{Robert} {et~al.}(2020){Robert}, {M{\'e}heut}, \&
  {M{\'e}nard}}]{Robert_2020}
{Robert}, C.~M.~T., {M{\'e}heut}, H., \& {M{\'e}nard}, F. 2020, \aap, 641,
  A128, \dodoi{10.1051/0004-6361/201937414}

\bibitem[{{Schr{\"a}pler} \& {Henning}(2004)}]{Schapler_2004}
{Schr{\"a}pler}, R., \& {Henning}, T. 2004, \apj, 614, 960,
  \dodoi{10.1086/423831}

\bibitem[{{Semenov} {et~al.}(2018){Semenov}, {Favre}, {Fedele}, {Guilloteau},
  {Teague}, {Henning}, {Dutrey}, {Chapillon}, {Hersant}, \&
  {Pi{\'e}tu}}]{Semenov_2018}
{Semenov}, D., {Favre}, C., {Fedele}, D., {et~al.} 2018, \aap, 617, A28,
  \dodoi{10.1051/0004-6361/201832980}

\bibitem[{{Sierra} \& {Lizano}(2020)}]{Sierra_2020}
{Sierra}, A., \& {Lizano}, S. 2020, \apj, 892, 136,
  \dodoi{10.3847/1538-4357/ab7d32}

\bibitem[{{Sierra} {et~al.}(2019){Sierra}, {Lizano}, {Mac{\'\i}as},
  {Carrasco-Gonz{\'a}lez}, {Osorio}, \& {Flock}}]{Sierra_2019}
{Sierra}, A., {Lizano}, S., {Mac{\'\i}as}, E., {et~al.} 2019, \apj, 876, 7,
  \dodoi{10.3847/1538-4357/ab1265}

\bibitem[{{Sierra} {et~al.}(2021){Sierra}, {P{\'e}rez}, {Zhang}, {Law},
  {Guzm{\'a}n}, {Qi}, {Bosman}, {{\"O}berg}, {Andrews}, {Long}, {Teague},
  {Booth}, {Walsh}, {Wilner}, {M{\'e}nard}, {Cataldi}, {Czekala}, {Bae},
  {Huang}, {Bergner}, {Ilee}, {Benisty}, {Le Gal}, {Loomis}, {Tsukagoshi},
  {Liu}, {Yamato}, \& {Aikawa}}]{Sierra_2021}
{Sierra}, A., {P{\'e}rez}, L.~M., {Zhang}, K., {et~al.} 2021, \apjs, 257, 14,
  \dodoi{10.3847/1538-4365/ac1431}

\bibitem[{{Strom} {et~al.}(1989){Strom}, {Strom}, {Edwards}, {Cabrit}, \&
  {Skrutskie}}]{Strom_1989}
{Strom}, K.~M., {Strom}, S.~E., {Edwards}, S., {Cabrit}, S., \& {Skrutskie},
  M.~F. 1989, \aj, 97, 1451, \dodoi{10.1086/115085}

\bibitem[{{Takeuchi} \& {Lin}(2002)}]{Takeuchi_2002}
{Takeuchi}, T., \& {Lin}, D.~N.~C. 2002, \apj, 581, 1344,
  \dodoi{10.1086/344437}

\bibitem[{{Tazaki} {et~al.}(2019){Tazaki}, {Tanaka}, {Kataoka}, {Okuzumi}, \&
  {Muto}}]{Tazaki_2019}
{Tazaki}, R., {Tanaka}, H., {Kataoka}, A., {Okuzumi}, S., \& {Muto}, T. 2019,
  \apj, 885, 52, \dodoi{10.3847/1538-4357/ab45f0}

\bibitem[{Tazzari(2017)}]{uvplot}
Tazzari, M. 2017, mtazzari/uvplot,  Zenodo, \dodoi{10.5281/zenodo.1003113}

\bibitem[{Teague(2019)}]{GoFish}
Teague, R. 2019, The Journal of Open Source Software, 4, 1632,
  \dodoi{10.21105/joss.01632}

\bibitem[{{Terada} {et~al.}(2023){Terada}, {Liu}, {Mkrtichian}, {Sai},
  {Konishi}, {Jiang}, {Muto}, {Hashimoto}, \& {Tamura}}]{Terada_2023}
{Terada}, Y., {Liu}, H.~B., {Mkrtichian}, D., {et~al.} 2023, \apj, 953, 147,
  \dodoi{10.3847/1538-4357/acdedf}

\bibitem[{{Testi} {et~al.}(2001){Testi}, {Natta}, {Shepherd}, \&
  {Wilner}}]{Testi_2001}
{Testi}, L., {Natta}, A., {Shepherd}, D.~S., \& {Wilner}, D.~J. 2001, \apj,
  554, 1087, \dodoi{10.1086/321406}

\bibitem[{{Testi} {et~al.}(2014){Testi}, {Birnstiel}, {Ricci}, {Andrews},
  {Blum}, {Carpenter}, {Dominik}, {Isella}, {Natta}, {Williams}, \&
  {Wilner}}]{Testi_2014}
{Testi}, L., {Birnstiel}, T., {Ricci}, L., {et~al.} 2014, in Protostars and
  Planets VI, ed. H.~{Beuther}, R.~S. {Klessen}, C.~P. {Dullemond}, \&
  T.~{Henning}, 339--361, \dodoi{10.2458/azu_uapress_9780816531240-ch015}

\bibitem[{{Thompson} {et~al.}(2017){Thompson}, {Moran}, \&
  {Swenson}}]{Thompson_2017}
{Thompson}, A.~R., {Moran}, J.~M., \& {Swenson}, George~W., J. 2017,
  {Interferometry and Synthesis in Radio Astronomy, 3rd Edition},
  \dodoi{10.1007/978-3-319-44431-4}

\bibitem[{{Toomre}(1964)}]{Toomre_1964}
{Toomre}, A. 1964, \apj, 139, 1217, \dodoi{10.1086/147861}

\bibitem[{{Ubeira Gabellini} {et~al.}(2019){Ubeira Gabellini}, {Miotello},
  {Facchini}, {Ragusa}, {Lodato}, {Testi}, {Benisty}, {Bruderer}, {T.
  Kurtovic}, {Andrews}, {Carpenter}, {Corder}, {Dipierro}, {Ercolano},
  {Fedele}, {Guidi}, {Henning}, {Isella}, {Kwon}, {Linz}, {McClure}, {Perez},
  {Ricci}, {Rosotti}, {Tazzari}, \& {Wilner}}]{Ubeira_2019}
{Ubeira Gabellini}, M.~G., {Miotello}, A., {Facchini}, S., {et~al.} 2019,
  \mnras, 486, 4638, \dodoi{10.1093/mnras/stz1138}

\bibitem[{{Ueda} {et~al.}(2020){Ueda}, {Kataoka}, \& {Tsukagoshi}}]{Ueda_2020}
{Ueda}, T., {Kataoka}, A., \& {Tsukagoshi}, T. 2020, \apj, 893, 125,
  \dodoi{10.3847/1538-4357/ab8223}

\bibitem[{{van der Marel}(2023)}]{vanderMarel_2023}
{van der Marel}, N. 2023, European Physical Journal Plus, 138, 225,
  \dodoi{10.1140/epjp/s13360-022-03628-0}

\bibitem[{{van der Marel} {et~al.}(2016){van der Marel}, {van Dishoeck},
  {Bruderer}, {Andrews}, {Pontoppidan}, {Herczeg}, {van Kempen}, \&
  {Miotello}}]{vanderMarel_2016}
{van der Marel}, N., {van Dishoeck}, E.~F., {Bruderer}, S., {et~al.} 2016,
  \aap, 585, A58, \dodoi{10.1051/0004-6361/201526988}

\bibitem[{{van der Marel} {et~al.}(2015){van der Marel}, {van Dishoeck},
  {Bruderer}, {P{\'e}rez}, \& {Isella}}]{vanderMarel_2015}
{van der Marel}, N., {van Dishoeck}, E.~F., {Bruderer}, S., {P{\'e}rez}, L., \&
  {Isella}, A. 2015, \aap, 579, A106, \dodoi{10.1051/0004-6361/201525658}

\bibitem[{{van der Marel} {et~al.}(2021){van der Marel}, {Birnstiel}, {Garufi},
  {Ragusa}, {Christiaens}, {Price}, {Sallum}, {Muley}, {Francis}, \&
  {Dong}}]{vanderMarel_2021}
{van der Marel}, N., {Birnstiel}, T., {Garufi}, A., {et~al.} 2021, \aj, 161,
  33, \dodoi{10.3847/1538-3881/abc3ba}

\bibitem[{{Varga} {et~al.}(2021){Varga}, {Hogerheijde}, {van Boekel},
  {Klarmann}, {Petrov}, {Waters}, {Lagarde}, {Pantin}, {Berio}, {Weigelt},
  {Robbe-Dubois}, {Lopez}, {Millour}, {Augereau}, {Meheut}, {Meilland},
  {Henning}, {Jaffe}, {Bettonvil}, {Bristow}, {Hofmann}, {Matter}, {Zins},
  {Wolf}, {Allouche}, {Donnan}, {Schertl}, {Dominik}, {Heininger}, {Lehmitz},
  {Cruzal{\`e}bes}, {Glindemann}, {Meisenheimer}, {Paladini}, {Sch{\"o}ller},
  {Woillez}, {Venema}, {Kokoulina}, {Yoffe}, {{\'A}brah{\'a}m}, {Abadie},
  {Abuter}, {Accardo}, {Adler}, {Ag{\'o}cs}, {Antonelli}, {B{\"o}hm}, {Bailet},
  {Bazin}, {Beckmann}, {Beltran}, {Boland}, {Bourget}, {Brast}, {Bresson},
  {Burtscher}, {Castillo}, {Chelli}, {Cid}, {Clausse}, {Connot}, {Conzelmann},
  {Danchi}, {De Haan}, {Delbo}, {Ebert}, {Elswijk}, {Fantei}, {Frahm},
  {G{\'a}mez Rosas}, {Gabasch}, {Gallenne}, {Garces}, {Girard}, {Gont{\'e}},
  {Gonz{\'a}lez Herrera}, {Graser}, {Guajardo}, {Guitton}, {Haubois}, {Hron},
  {Hubin}, {Huerta}, {Isbell}, {Ives}, {Jakob}, {Jask{\'o}}, {Jochum}, {Klein},
  {Kragt}, {Kroes}, {Kuindersma}, {Labadie}, {Laun}, {Le Poole}, {Leinert},
  {Lizon}, {Lopez}, {M{\'e}rand}, {Marcotto}, {Mauclert}, {Maurer}, {Mehrgan},
  {Meisner}, {Meixner}, {Mellein}, {Mohr}, {Morel}, {Mosoni}, {Navarro},
  {Neumann}, {Nu{\ss}baum}, {Pallanca}, {Pasquini}, {Percheron}, {Pott},
  {Pozna}, {Ridinger}, {Rigal}, {Riquelme}, {Rivinius}, {Roelfsema}, {Rohloff},
  {Rousseau}, {Schuhler}, {Schuil}, {Soulain}, {Stee}, {Stephan}, {ter Horst},
  {Tromp}, {Vakili}, {van Duin}, {Vinther}, {Wittkowski}, \&
  {Wrhel}}]{Varga_2021}
{Varga}, J., {Hogerheijde}, M., {van Boekel}, R., {et~al.} 2021, \aap, 647,
  A56, \dodoi{10.1051/0004-6361/202039400}

\bibitem[{{Wahhaj} {et~al.}(2010){Wahhaj}, {Cieza}, {Koerner}, {Stapelfeldt},
  {Padgett}, {Case}, {Keller}, {Mer{\'\i}n}, {Evans}, {Harvey}, {Sargent}, {van
  Dishoeck}, {Allen}, {Blake}, {Brooke}, {Chapman}, {Mundy}, \&
  {Myers}}]{Wahhaj_2010}
{Wahhaj}, Z., {Cieza}, L., {Koerner}, D.~W., {et~al.} 2010, \apj, 724, 835,
  \dodoi{10.1088/0004-637X/724/2/835}

\bibitem[{{Weidenschilling}(1977)}]{Weidenschilling_1977}
{Weidenschilling}, S.~J. 1977, \mnras, 180, 57, \dodoi{10.1093/mnras/180.2.57}

\bibitem[{{Whipple}(1972)}]{Whipple_1972}
{Whipple}, F.~L. 1972, in From Plasma to Planet, ed. A.~{Elvius}, 211

\bibitem[{{W{\"o}lfer} {et~al.}(2023){W{\"o}lfer}, {Facchini}, {van der Marel},
  {van Dishoeck}, {Benisty}, {Bohn}, {Francis}, {Izquierdo}, \&
  {Teague}}]{Wolfer_2023}
{W{\"o}lfer}, L., {Facchini}, S., {van der Marel}, N., {et~al.} 2023, \aap,
  670, A154, \dodoi{10.1051/0004-6361/202243601}

\bibitem[{{Yang} {et~al.}(2024){Yang}, {Fern{\'a}ndez-L{\'o}pez}, {Li},
  {Stephens}, {Looney}, {Lin}, \& {Harrison}}]{Yang_2024}
{Yang}, H., {Fern{\'a}ndez-L{\'o}pez}, M., {Li}, Z.-Y., {et~al.} 2024, \apj,
  963, 134, \dodoi{10.3847/1538-4357/ad2346}

\bibitem[{{Yang} {et~al.}(2016){Yang}, {Li}, {Looney}, \&
  {Stephens}}]{Yang_2016}
{Yang}, H., {Li}, Z.-Y., {Looney}, L., \& {Stephens}, I. 2016, \mnras, 456,
  2794, \dodoi{10.1093/mnras/stv2633}

\bibitem[{{Zapata} {et~al.}(2017){Zapata}, {Rodr{\'\i}guez}, \&
  {Palau}}]{Zapata_2017}
{Zapata}, L.~A., {Rodr{\'\i}guez}, L.~F., \& {Palau}, A. 2017, \apj, 834, 138,
  \dodoi{10.3847/1538-4357/834/2/138}

\bibitem[{{Zhang} {et~al.}(2023){Zhang}, {Zhu}, {Ueda}, {Kataoka}, {Sierra},
  {Carrasco-Gonz{\'a}lez}, \& {Mac{\'\i}as}}]{Zhang_2023}
{Zhang}, S., {Zhu}, Z., {Ueda}, T., {et~al.} 2023, \apj, 953, 96,
  \dodoi{10.3847/1538-4357/acdb4e}

\bibitem[{{Zhu} {et~al.}(2019){Zhu}, {Zhang}, {Jiang}, {Kataoka}, {Birnstiel},
  {Dullemond}, {Andrews}, {Huang}, {P{\'e}rez}, {Carpenter}, {Bai}, {Wilner},
  \& {Ricci}}]{Zhu_2019}
{Zhu}, Z., {Zhang}, S., {Jiang}, Y.-F., {et~al.} 2019, \apjl, 877, L18,
  \dodoi{10.3847/2041-8213/ab1f8c}

\bibitem[{{Zurlo} {et~al.}(2020){Zurlo}, {Cugno}, {Montesinos}, {Perez},
  {Canovas}, {Casassus}, {Christiaens}, {Cieza}, \& {Huelamo}}]{Zurlo_2020}
{Zurlo}, A., {Cugno}, G., {Montesinos}, M., {et~al.} 2020, \aap, 633, A119,
  \dodoi{10.1051/0004-6361/201936891}

\end{thebibliography}
\bibliographystyle{aasjournal}

\end{document}